\documentclass[prb,aps,noshowpacs,preprintnumbers,amsmath,amssymb,twocolumn,superscriptaddress]{revtex4-2}
\usepackage{graphicx}
\usepackage{dcolumn}
\usepackage{epsfig}
\usepackage{bbold,bm}
\usepackage[export]{adjustbox}
\usepackage[normalem]{ulem}
\usepackage{graphicx}
\usepackage{mathtools}
\usepackage{hyperref}
\usepackage{xcolor}

\def\multiset#1#2{\ensuremath{\left(\kern-.3em\left(\genfrac{}{}{0pt}{}{#1}{#2}\right)\kern-.3em\right)}}

\def\ket#1{\mathinner{|{#1}\rangle}}

\def\tr{\mathrm{Tr}}

\makeatother

\begin{document}


\title{Classical and Quantum Chaos in Chirally-Driven, Dissipative Bose-Hubbard Systems}


\author{Daniel Dahan}
\affiliation{Department of Physics, Ben-Gurion University of the Negev, Beer-Sheva
	8410501, Israel}

\author{Geva Arwas}
\affiliation{Department of Physics of Complex Systems,
	Weizmann Institute of Science, Rehovot, 76100, Israel}

\author{Eytan Grosfeld}
\affiliation{Department of Physics, Ben-Gurion University of the Negev, Beer-Sheva
	8410501, Israel}


\begin{abstract}
We study the dissipative Bose-Hubbard model on a small ring of sites in the presence of a chiral drive and explore its long-time dynamical structure using the mean field equations and by simulating the quantum master equation. Remarkably, for large enough drivings, we find that the system admits, in a wide range of parameters, a chaotic attractor at the mean-field level, which manifests as a complex Wigner function on the quantum level. The latter is shown to have the largest weight around the approximate region of phase space occupied by the chaotic attractor. We demonstrate that this behavior could be revealed via measurement of various bosonic correlation functions. In particular, we employ open system methods to calculate the out-of-time-ordered correlator, whose exponential growth signifies a positive quantum Lyapunov exponent in our system. This can open a pathway to the study of chaotic dynamics in interacting systems of photons.
\end{abstract}

\date{\today}

\maketitle


\emph{Introduction.}--- Methods to confine and control photons using solid state materials have opened a pathway to the generation of novel liquids of light \cite{carusotto2013quantum}, exhibiting a wide range of phenomena, from superfluidity \cite{amo2009superfluidity} to topological states \cite{ozawa2019topological}. Coupled cavity arrays evolved into a valuable experimental framework for the study of lattices of photons in the presence of interactions, drive and dissipation \cite{deng2002condensation,schmidt2013circuit,fitzpatrick2017observation}. Such systems benefit from the exquisite fabrication techniques of solid state materials as well as from the high coherence properties of the photons, making them an attractive venue for future applications~\cite{amo2016exciton}.

Even a small number of driven coupled cavities can exhibit rich behavior: a single cavity gives rise to regions of bistability and to critical slowing down~\cite{fink2018signatures}; cavity dimers demonstrate Josephson oscillations~\cite{abbarchi2013macroscopic} and parametric instabilities \cite{zambon2020parametric} and could be driven into a more exotic quantum limit-cycle if the dissipation is further engineered~\cite{lledo2019driven}; the levels of a six cavity ring were demonstrated to be individually excitable, generating chirality, with their population exhibiting hysteretic behavior, indicative of a bistability~\cite{sala2015spin}; and many more. These phenomena explore at most a quasi-two-dimensional manifold of phase space. Yet, on the classical, mean-field level such coupled cavities are described as driven-dissipative coupled nonlinear oscillators, that are in principle capable of exhibiting chaotic behavior~\cite{solnyshkov2009chaotic,gavrilov2016towards,sanchez2020autonomous}. However, it remains unclear what are the key ingredients which are required to generate chaotic behavior in such systems. Further, once the interactions increase and the system crosses into a more quantum regime, what implications does the emergence of classical chaos carry for the quantum behaviour of the system?

\begin{figure}[b]
\includegraphics[width=0.6\linewidth]{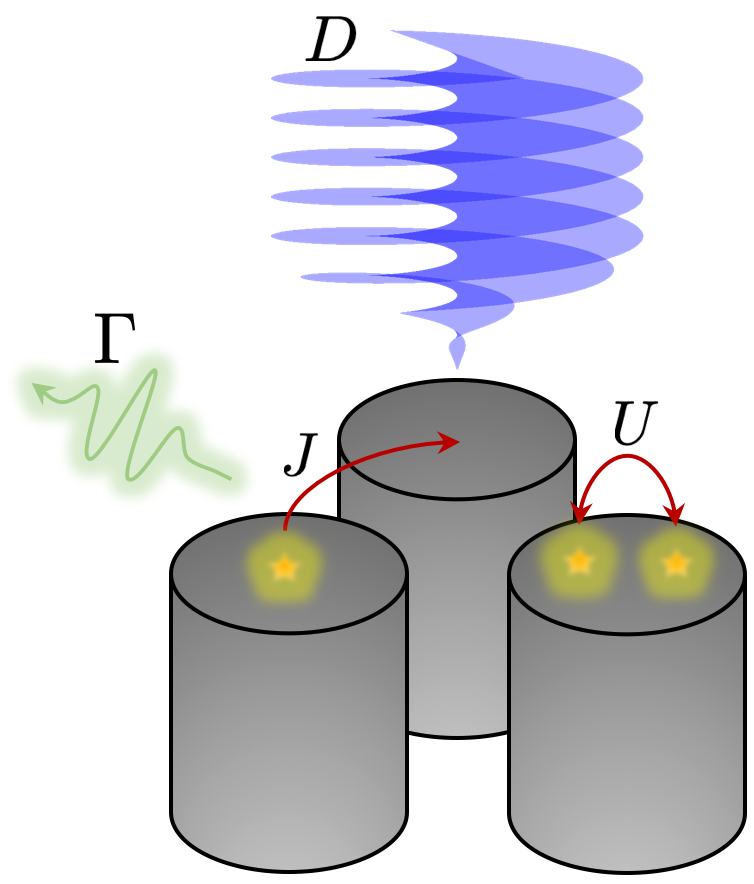}
	\caption{
		{\bf Illustration of the main setup considered in the text.} A ring of $M=3$ coupled cavities interacting with chiral light, with hopping $J$, on-cavity interaction $U$, dissipation $\Gamma$ and chiral drive $D$.}
	\label{fig:setup} 
	\vspace{-5mm}
\end{figure}

\begin{figure*}[t]
	\begin{minipage}[t]{0.271\linewidth}
		\includegraphics[width=\linewidth]{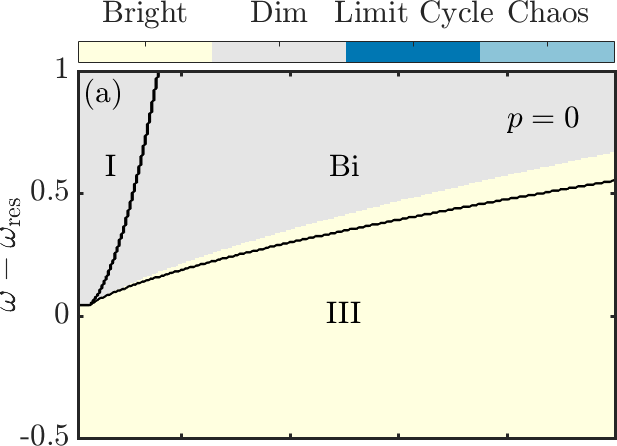}\vspace{0.7mm}
		\includegraphics[width=\linewidth]{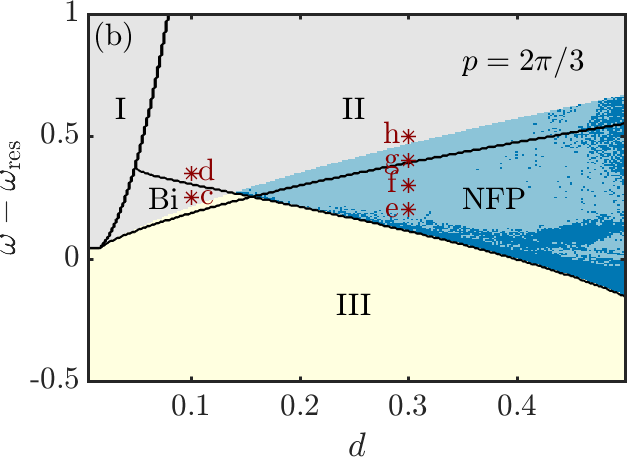}
	\end{minipage}
\hspace{0.1cm}
	\begin{minipage}[t]{0.55\linewidth}
		\begin{minipage}[t]{\linewidth}
			\includegraphics[scale=0.25]{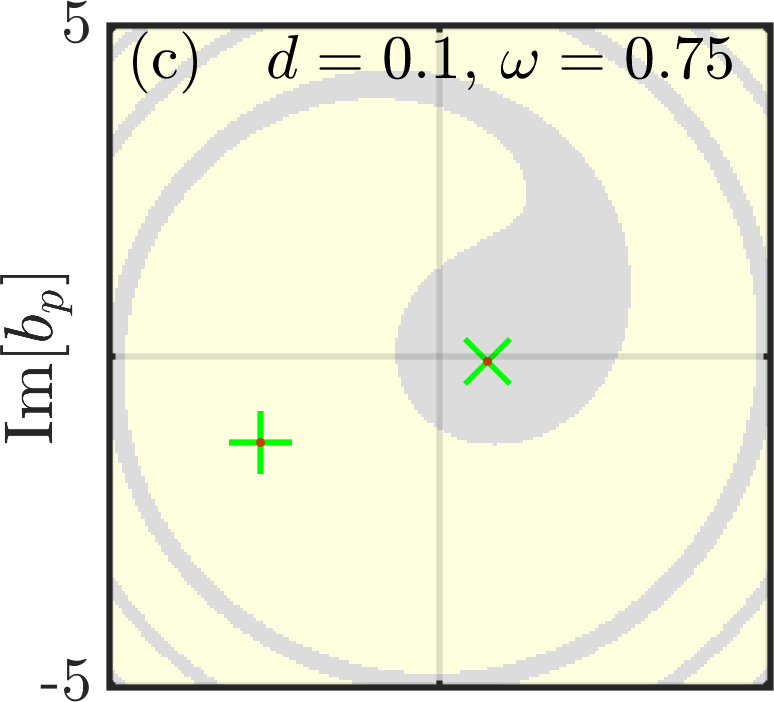}
			\hfill
			\includegraphics[scale=0.25]{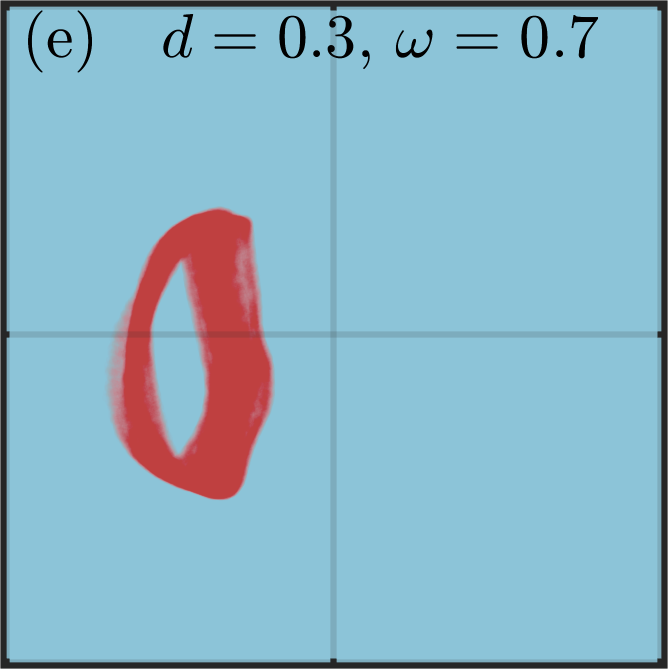}
			\hfill
			\includegraphics[scale=0.25]{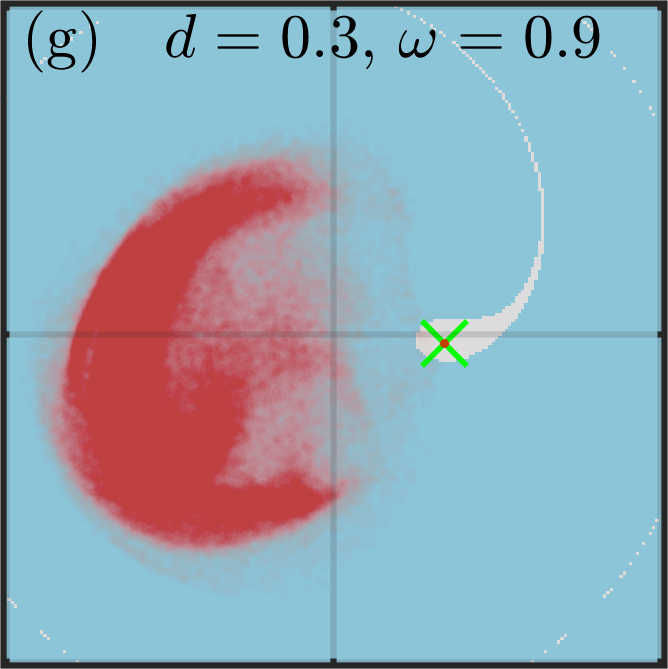}	
		\end{minipage}\vspace{0.7mm}
		\begin{minipage}[b]{\linewidth}
			\includegraphics [scale=0.25]{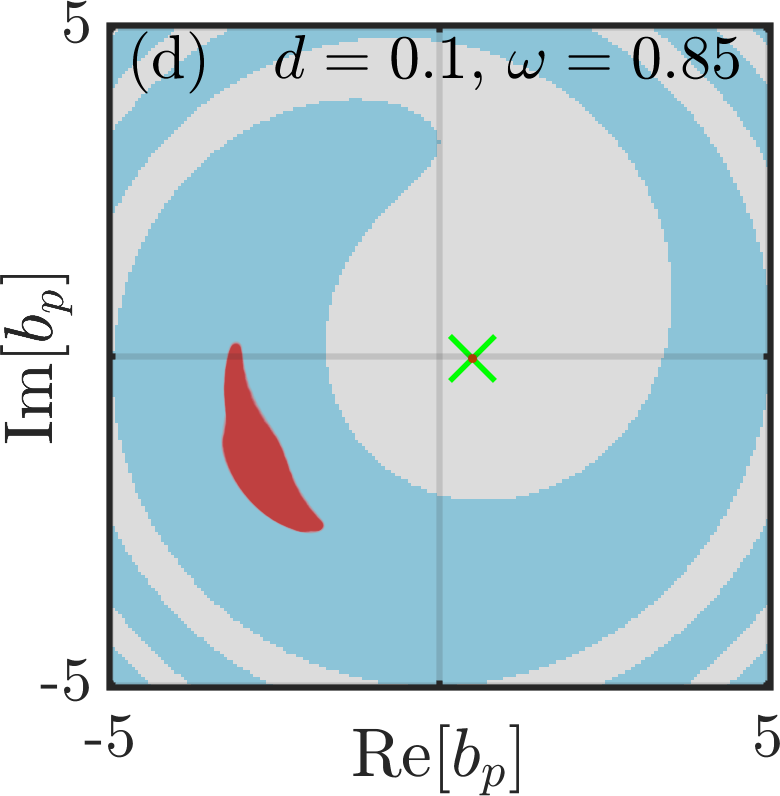}
			\hfill
			\includegraphics[scale=0.25]{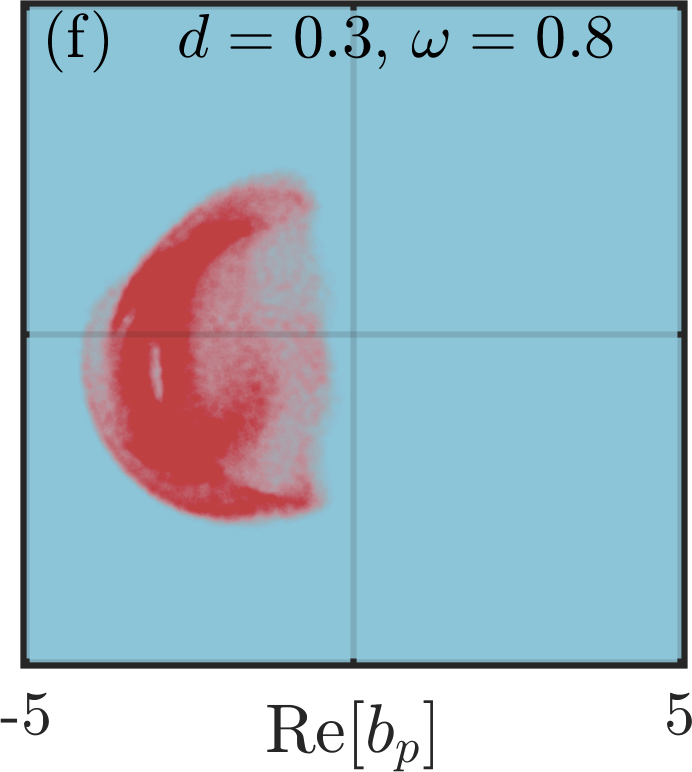}
			\hfill
			\includegraphics [scale=0.25]{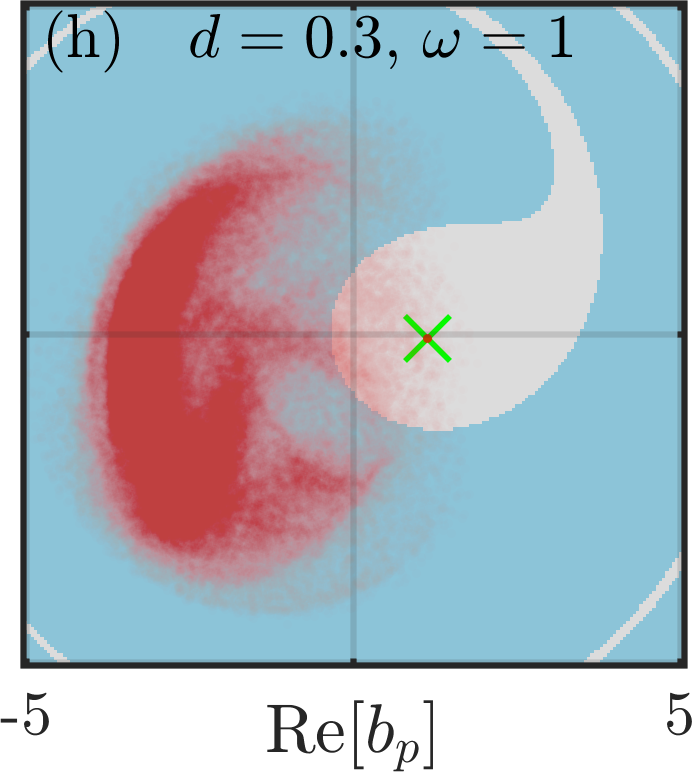}   
		\end{minipage}
	\end{minipage}
	\caption{
		{\bf Mean-field phase diagram and classical basins of attraction} for $M=3$ cavities on a ring. (a) The phase diagram for zero momentum drive, as function of the strength of the drive ($d$) and its frequency ($\omega$)  with respect to the non-interacting resonant frequency ($\omega_{\text{res}}=-\cos p$), displaying the bright (III) and dim (I and II) regimes and a bistability (Bi) regime (the subregions I, II are further detailed in \cite{appmat}). (b) The same for finite momentum ($p= 2\pi/3$) drive, giving rise to a region where no fixed point (NFP) is stable at larger drivings. The right panels visualize the basins of attraction of the different attractors for a finite chiral drive: the chaotic basins (light blue), the bright fixed point basin (yellow) and the dim fixed point basin (gray) [see the red asteriks in (b) for the corresponding parameters]. The ending points of the sampled trajectory (for $Jt_\text{m}=2000$) are projected onto the resonant plane (red).  The green $\times$ ($+$) markers correspond to the dim (bright) stable fixed point solution. Here $u=0.1$ and $\gamma=0.05$.}
	\label{fig:phase-diagram} 
	\vspace{-5mm}
\end{figure*}

In this paper we consider a ring of three coupled cavities, driven by finite orbital momentum light \cite{dholakia1996second,martinelli2004orbital,zambon2019circular} (illustrated in Fig.~\ref{fig:setup}). We demonstrate that as a direct result of the chirality of the applied drive, the system exhibits robust chaotic dynamics over a large part of its mean-field phase diagram (see Fig.~\ref{fig:phase-diagram}b in comparison to Fig.~\ref{fig:phase-diagram}a). On this level, the chaos is manifested as a dense, compact attractor in phase space. For the quantum problem, a full quantum trajectory approach \cite{molmer1993monte} reveals a complex Wigner function, indicative of the underlying chaos. The $g^{(1)}$ correlator places the system on the chaotic attractor, while the $g^{(2)}$ correlator reveals signatures of the chaotic behavior. We find that, most prominently, the out-of-time-ordered correlator (OTOC) reveals the onset of chaos via an exponential sensitivity to the application of a weak perturbation. These predictions could pave the way to the study of the interplay of classical and quantum chaotic dynamics and their properties in coupled cavity arrays of photons.

\emph{Proposed setup and model.}--- We consider the Bose-Hubbard Hamiltonian for $M$ sites (cavities) on a ring in the presence of drive and dissipation, with a driving term that is coherent with respect to both frequency and momentum,
\begin{align} \label{eq:hamiltonian}
	\hat{\mathcal H}=&  
	\frac{U}{2}\sum_{j=1}^M
	\hat{b}^\dagger_j \hat{b}^\dagger_j \hat{b}_j \hat{b}_j
	-\frac{J}{2} \sum_{j=1}^{M-1}\left(\hat{b}^\dagger_{j}\hat{b}_{j+1}+\hat{b}^\dagger_{j+1}\hat{b}_{j}\right) \nonumber \\ 
	&-\Omega\sum_{j=1}^M \hat{b}^\dagger_j \hat{b}_j +D \sum_{j=1}^M\left(e^{ip x_j}\hat{b}_j +e^{-i p x_j}\hat{b}^\dagger_j \right).
\end{align}
Here $\hat{b}_j^\dagger$ ($\hat{b}_j$) creates (annihilates) a boson at site $j$ ($j\in \{1,\ldots,M\}$ where $M$ is the number of sites), $x_j=j$ is the position of the site (choosing the units such that the lattice spacing $a=1$), $U>0$ is the on-site interaction strength, $J$ is the tunnelling amplitude and $D$, $\Omega$ and $p=\frac{2\pi}{M}m$ ($m\in\{0,\ldots,M-1\}$) are the amplitude, detuning and momentum of the drive. The Lindblad master equation governs the dynamics of the density matrix $\hat{\rho}$ in the presence of a boson loss rate $\Gamma$
\begin{align} \label{eq:lindblad}
	i \partial_t \hat{\rho} = [\hat{\mathcal H},\hat{\rho}]  + \frac{\Gamma}{2} \sum_j ( 2 \hat{b}_j \hat{\rho} \hat{b}^\dagger_j - \hat{b}^\dagger _j \hat{b}_j \hat{\rho} -\hat{\rho} \hat{b}^\dagger_j \hat{b}_j ).
\end{align}
It prescribes the non-Hermitian effective Hamiltonian $\hat{\mathcal H}_{\text{eff}}=\hat{\mathcal{H}}-i\frac{\Gamma}{2}\sum_{j=1}^M \hat{b}_j^\dagger \hat{b}_j$ which is the deterministic part of the Lindblad evolution; as well as the so-called ``recycling'' term, $\Gamma\sum_j \hat{b}_j\hat{\rho} \hat{b}_j^\dagger$. In the following we use the rescaled parameters $u=U/J$, $d=D/J$, $\omega=\Omega/J$ and $\gamma=\Gamma/J$.

\begin{figure*}[t]
	\begin{tabular}{ lll }
		\includegraphics[scale=0.3]{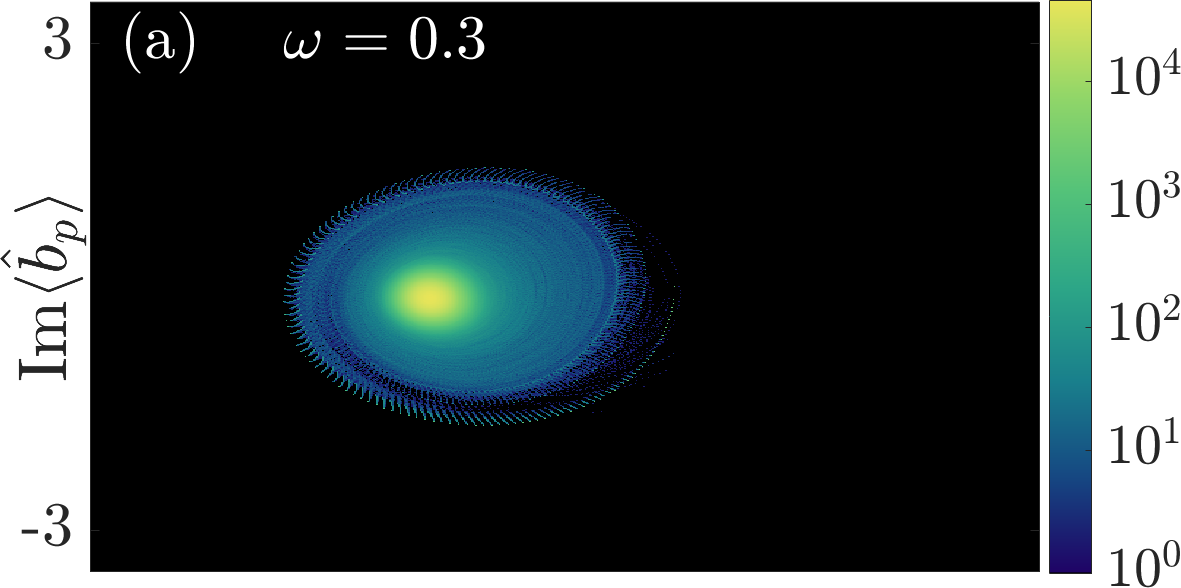}&
		\includegraphics[scale=0.3]{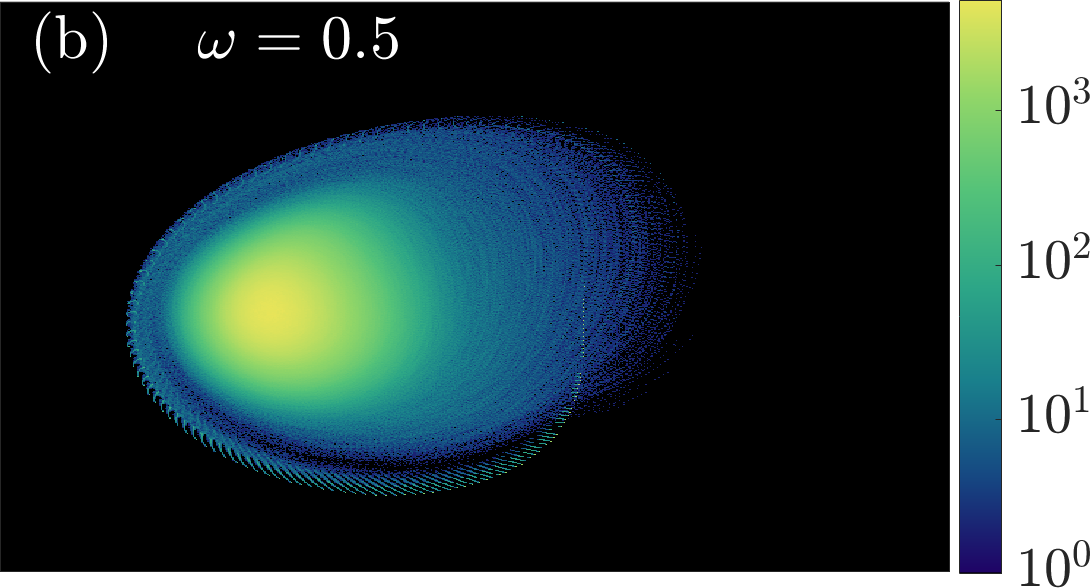}&
		\includegraphics[scale=0.3]{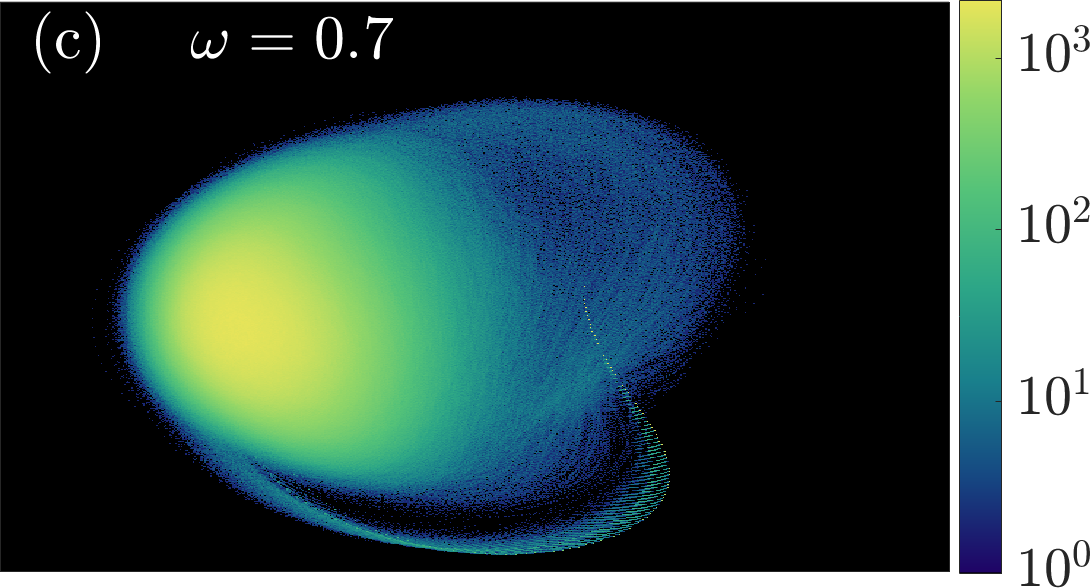}\\
		\includegraphics[scale=0.3]{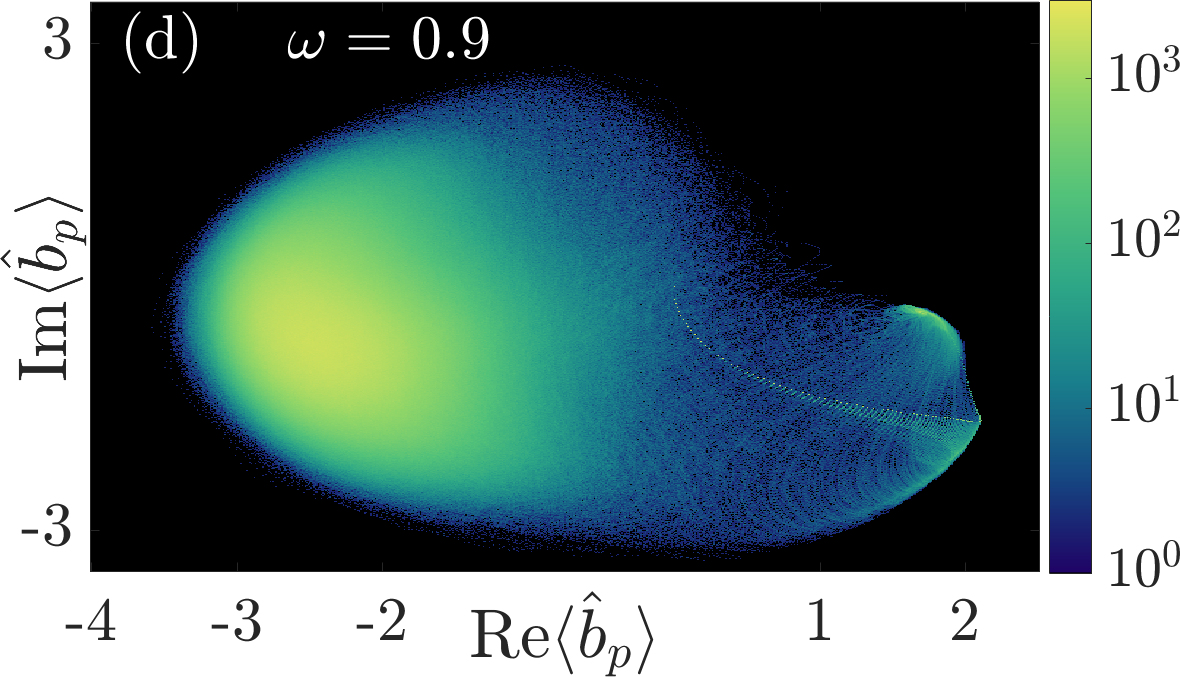}&
		\hspace{-0.8mm}\includegraphics[scale=0.3]{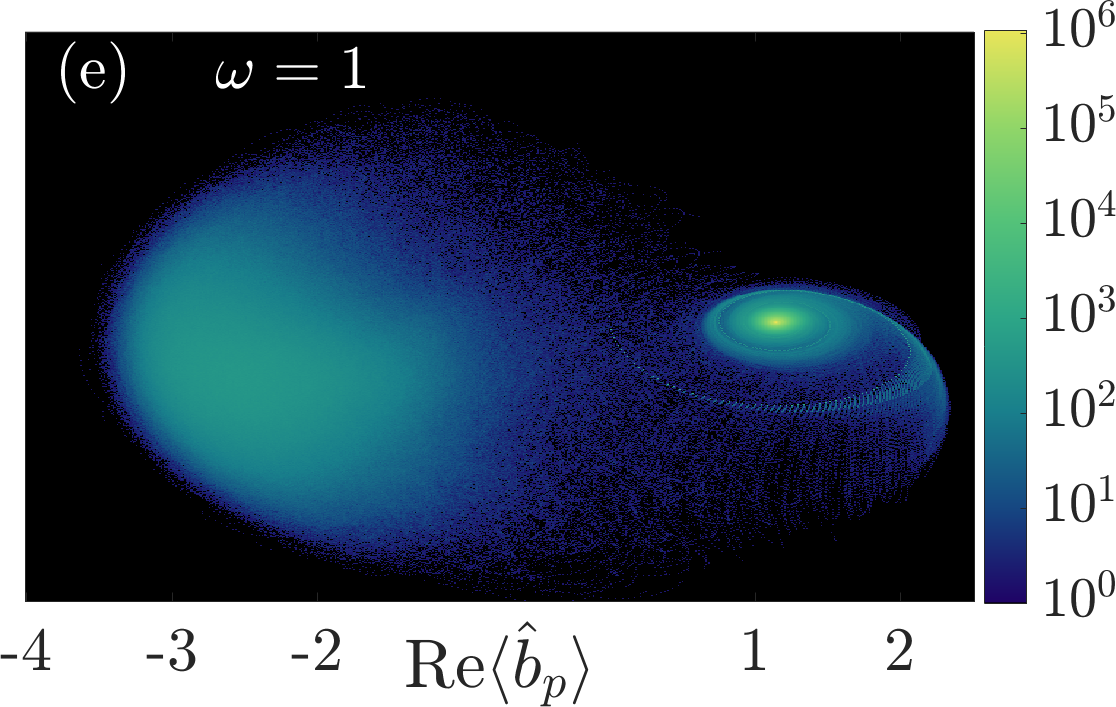}&
		\hspace{-0.8mm}\includegraphics[scale=0.3]{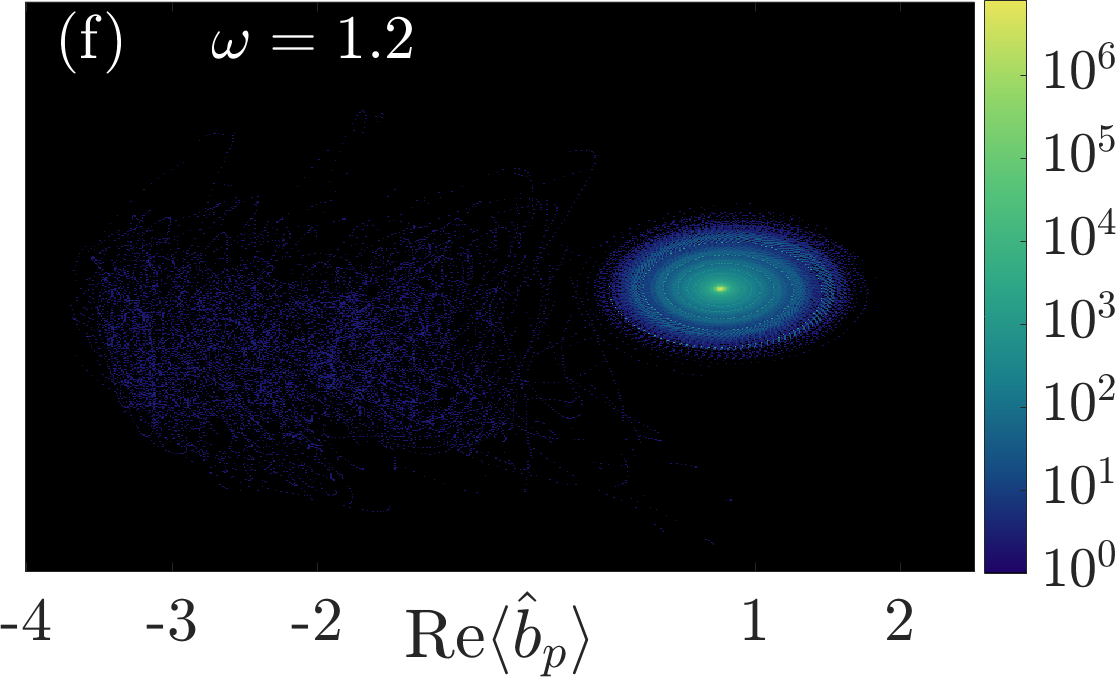}
	\end{tabular}
	\caption{
		{\bf Visualization of the quantum trajectories}, using a histogram, presented in log scale, with the quantum trajectories projected to the resonant plane. The evolution and interplay of the various fixed points are clearly visible as the frequency of the drive increases, including the emergence of the chaotic attractor replacing the bright fixed point and the nucleation of the dim fixed point at higher frequencies.
		Here $u=0.1$, $\gamma=0.05$, and $d=0.3$.}
	\label{fig:quantum_trajectories} 
	\vspace{-5mm}
\end{figure*}


\emph{Mean field phase diagram.}--- We start by analysing the mean-field equations of motion, derived from the hamiltonian $\hat{\mathcal{H}}_{\text{eff}}$. We analyze the fixed points using the ansatz that only a single mode participates in the dynamics, $\hat{b}_p=\frac{1}{\sqrt{M}}\sum_j e^{i p x_j}\hat{b}_j$, with the operators replaced by c-numbers. The regimes of existence and stability of the resulting fixed points are delineated by black lines in Fig.~\ref{fig:phase-diagram}a,b. These include ``dim'' and ``bright'' regimes which are each dominated by the effect of a single fixed point with low and high bosonic occupation, respectively.

The difference between a uniform and a chiral drive is already exposed in this mean-field level. These two types of drive lead to the two dramatically different phase diagrams, presented in Fig.~\ref{fig:phase-diagram}, with panel (a) describing the uniform case and (b) the chiral case.  In between the dim and bright regimes, for the uniform drive and for the chiral drive at low amplitudes, there exists a bistability regime (denoted ``Bi'') of the dim and bright fixed points. This picture changes considerably for the chiral drive at higher amplitudes, where a large regime where no fixed point (denoted ``NFP'') remains stable appears in the phase diagram.

To gain further intuition into the nature of this regime, we note that the actual steady state the system attains is decided by the initial condition. When we start from the vacuum state $|0\rangle$ with zero bosonic occupation, the resulting steady state is depicted by the shading in Fig.~\ref{fig:phase-diagram}a,b. In particular, in most of the bistable regime the system reaches the dim fixed point. In the region NFP it turns out that the system reaches no particular fixed point, but the long time dynamics remains confined in a certain dense region of phase space. In order to analyze it, we have to go beyond the single mode ansatz~\cite{appmat}.

We next relax the requirement that we start from the $|0\rangle$ state and consider the two-dimensional phase plane of initial conditions associated with $b_p$. Each initial condition is color-coded according to the attractor the system eventually reaches. The result is described in Fig.~\ref{fig:phase-diagram}c-h, and defines the basins of attraction of the relevant attractors. In addition, the long time trajectories of $b_p$ are superimposed in red.

The spiral shape of the basins was highlighted in \cite{kolovsky2020bistability} for the case of a single cavity. Our case is richer since the higher dimensional phase space allows for the possibility of classical chaos, which in our system is accessible due to the operation of a chiral drive. The onset of chaos is marked by an exponential sensitivity to initial conditions and is characterized by a positive (classical) Lyapunov exponent. For the NFP regime, we find that the system ends up in a strange attractor (SA), depicted by the red shape in Fig.~\ref{fig:phase-diagram}e-h. In fact the SA also exists beyond the region marked by NFP, and when the vacuum is contained in its basin of attraction the system will flow into this steady state, see e.g., Fig.~\ref{fig:phase-diagram}d. 

Once the interaction in the system increases the system will characteristically hold lower bosonic occupations and can cross into the quantum regime. In the next section we will investigate the quantum behavior of the system within and in the vicinity of the NFP region, and explore its properties.

\begin{figure*}[t]
	\hspace*{-0.7cm}		
	\begin{tabular}{ lllll }
		\includegraphics[scale=0.2,valign=m]{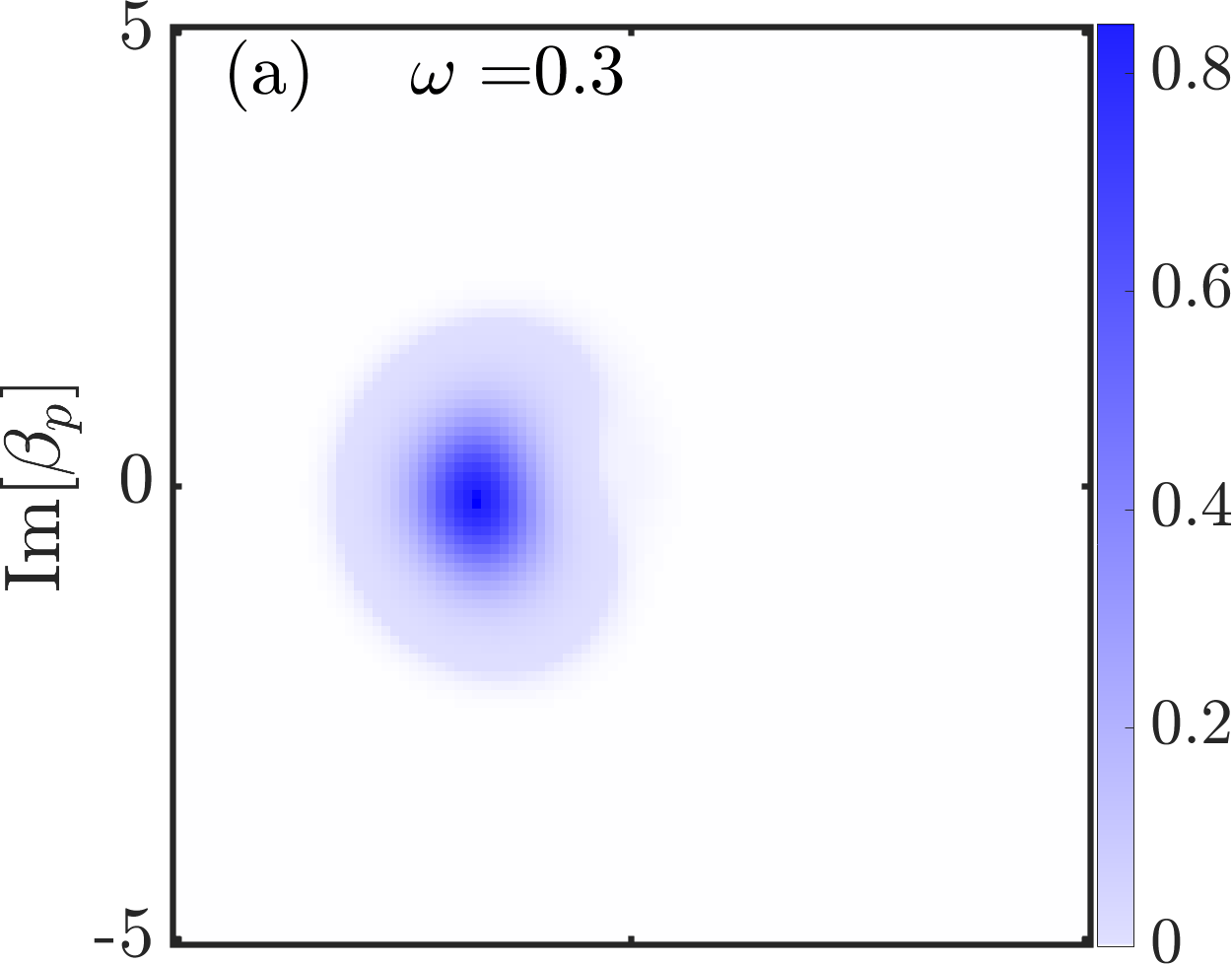} &       
		\hspace*{-1.1mm}\includegraphics[scale=0.2,valign=m]{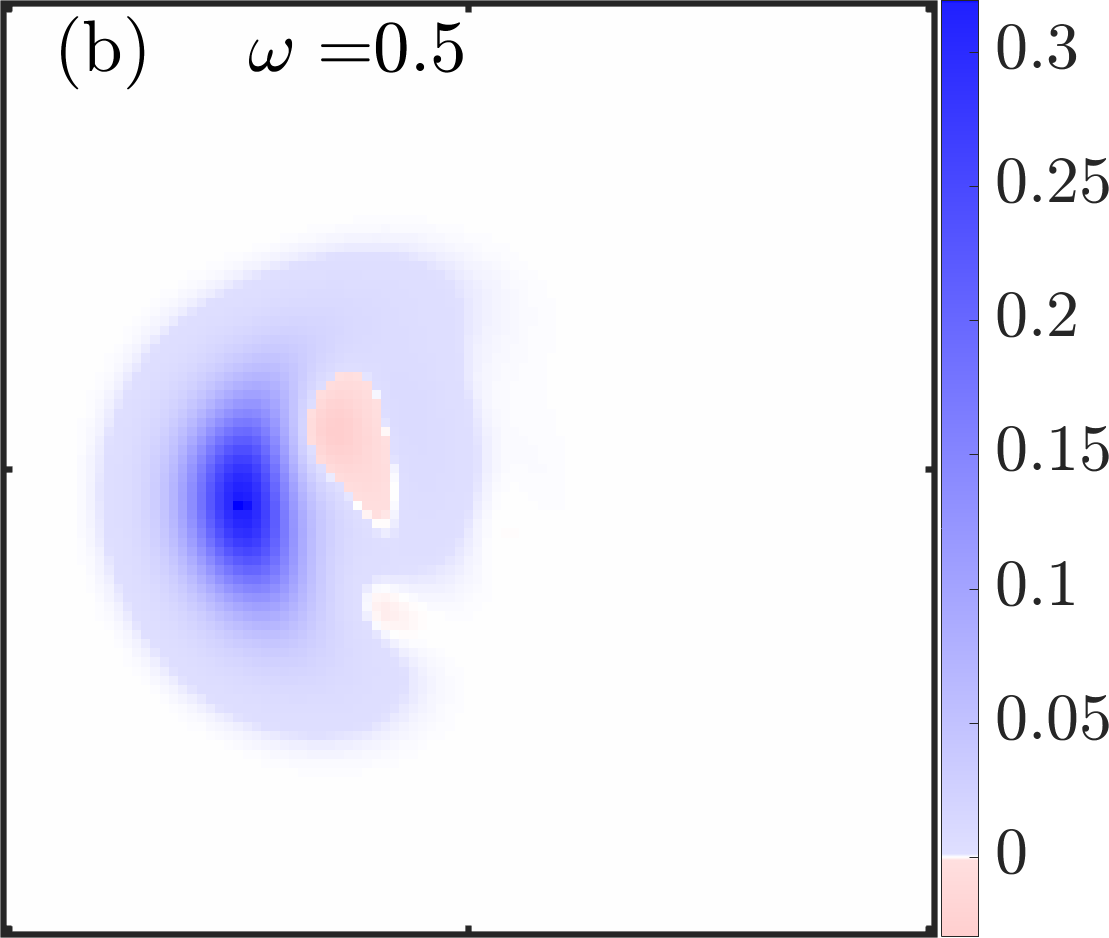} &
		\hspace*{-1.1mm}\includegraphics[scale=0.2,valign=m]{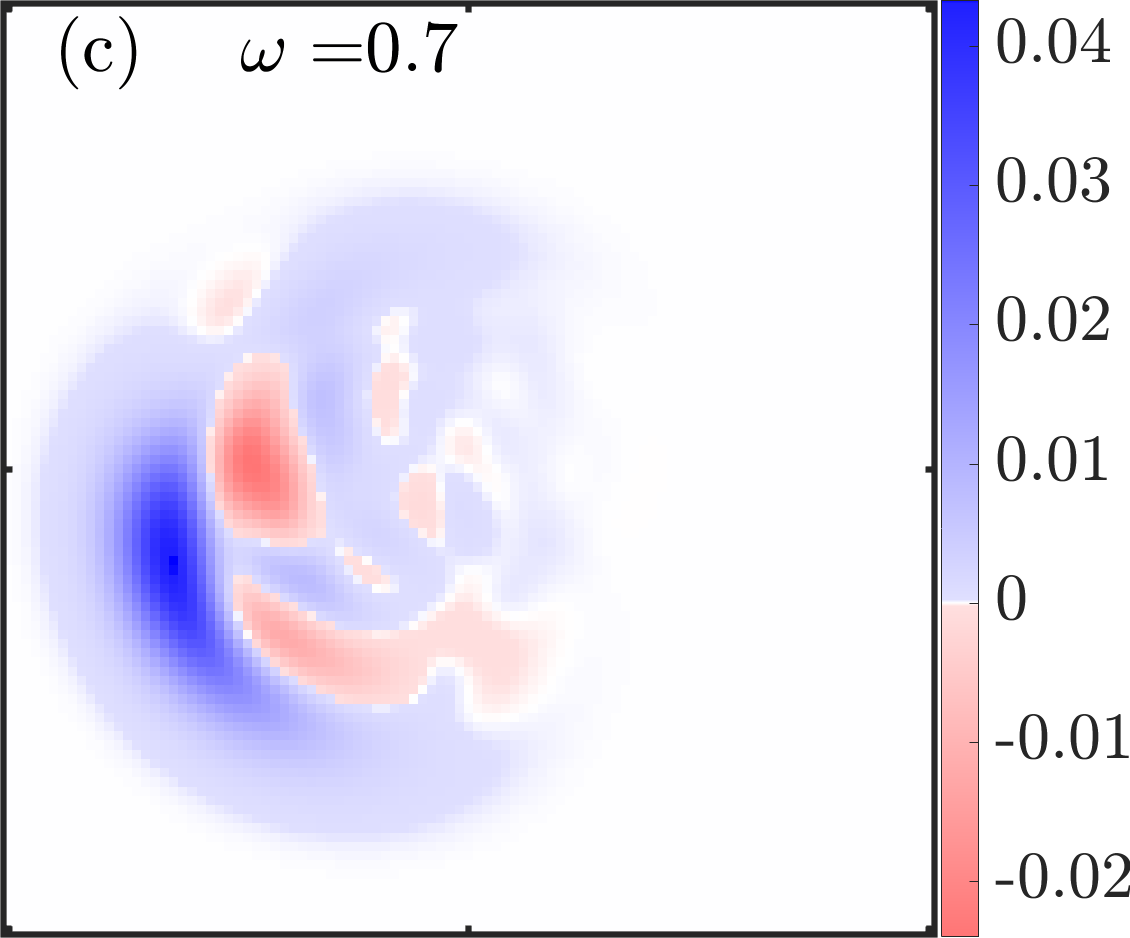} &
		\hspace*{-1.1mm}\includegraphics[scale=0.2,valign=m]{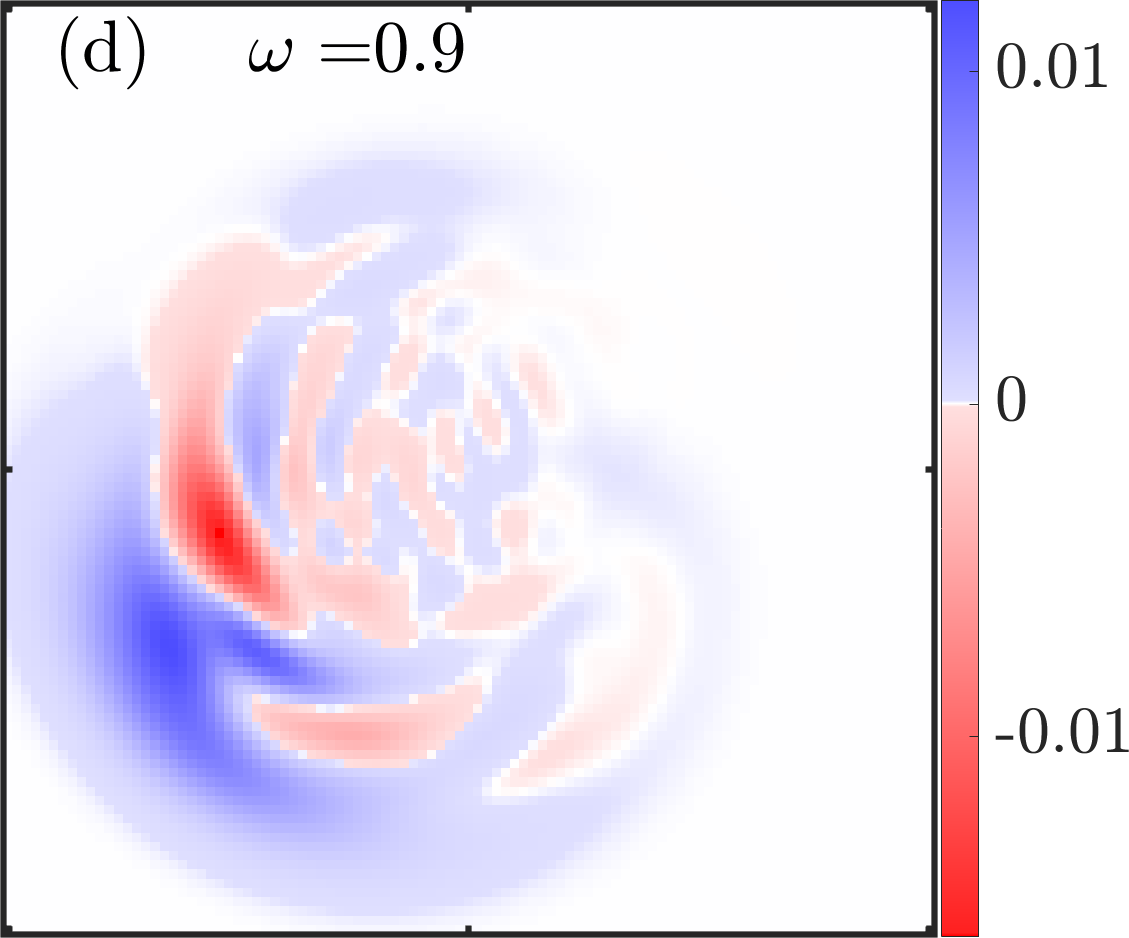} &
		\hspace*{-1.1mm}\includegraphics[scale=0.2,valign=m]{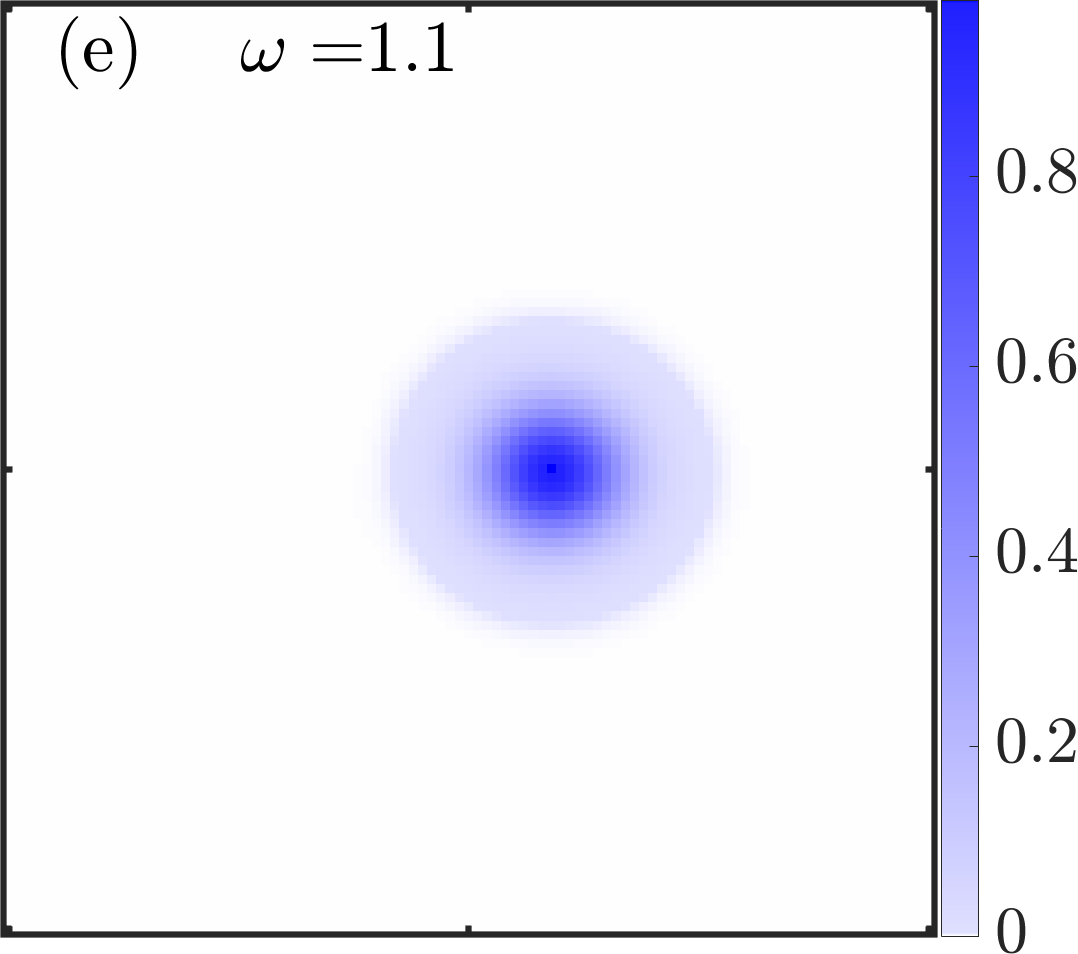} \\ \vspace{-0.3cm}
		\\
		\includegraphics[scale=0.2,valign=m]{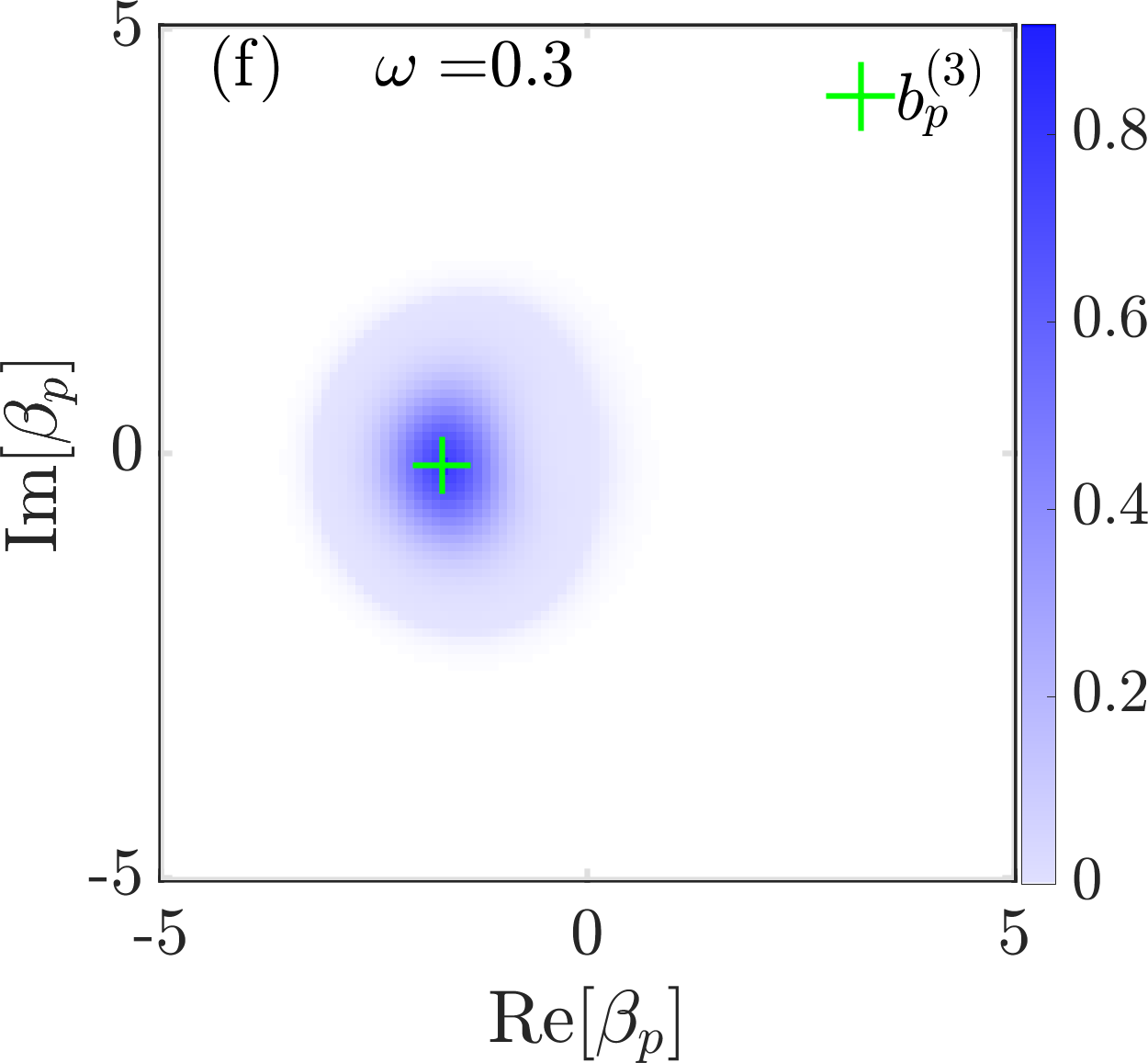} & 		 \hspace{-1.8mm}\includegraphics[scale=0.2,valign=m]{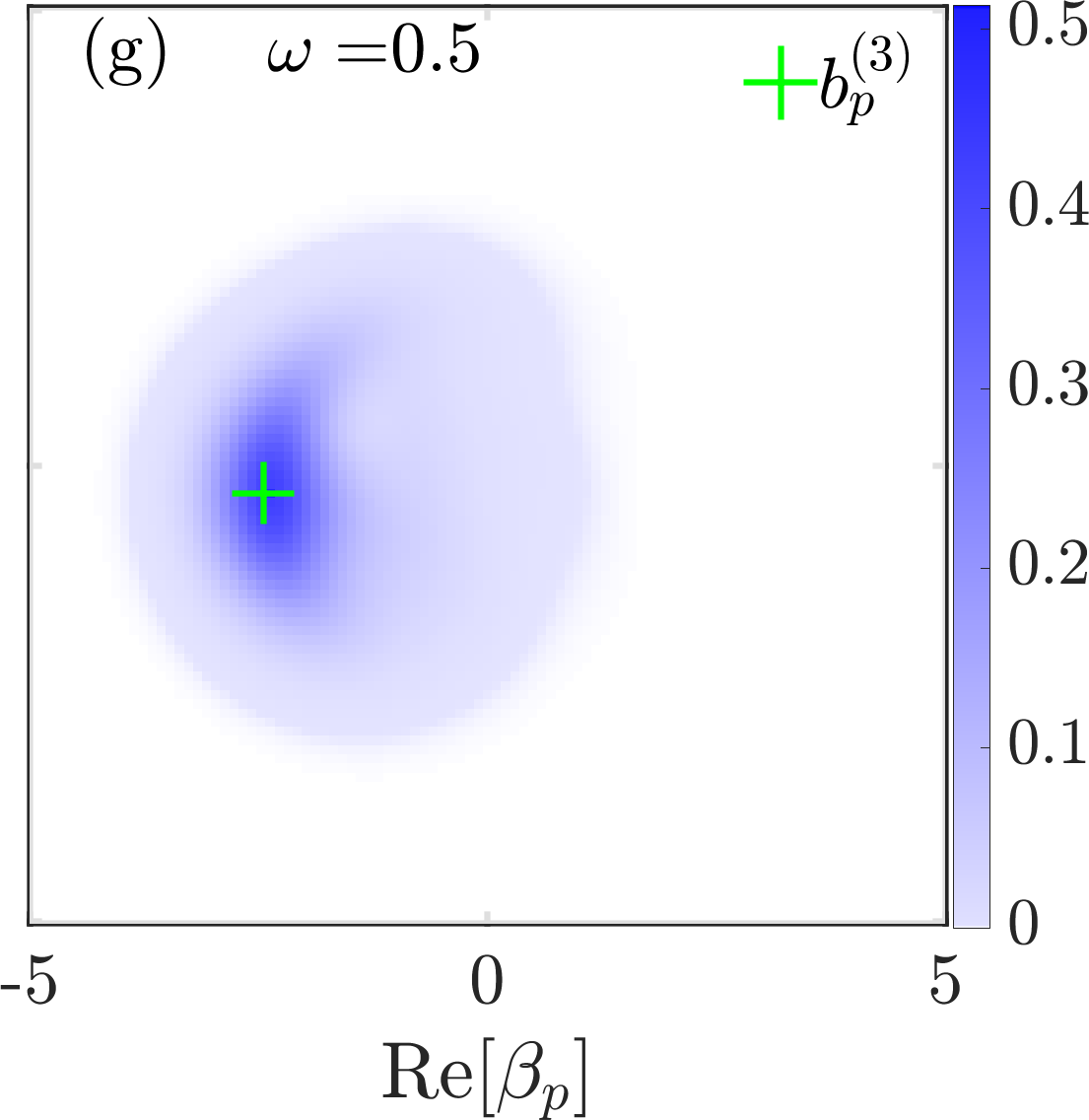} & \hspace{-1.8mm}\includegraphics[scale=0.2,valign=m]{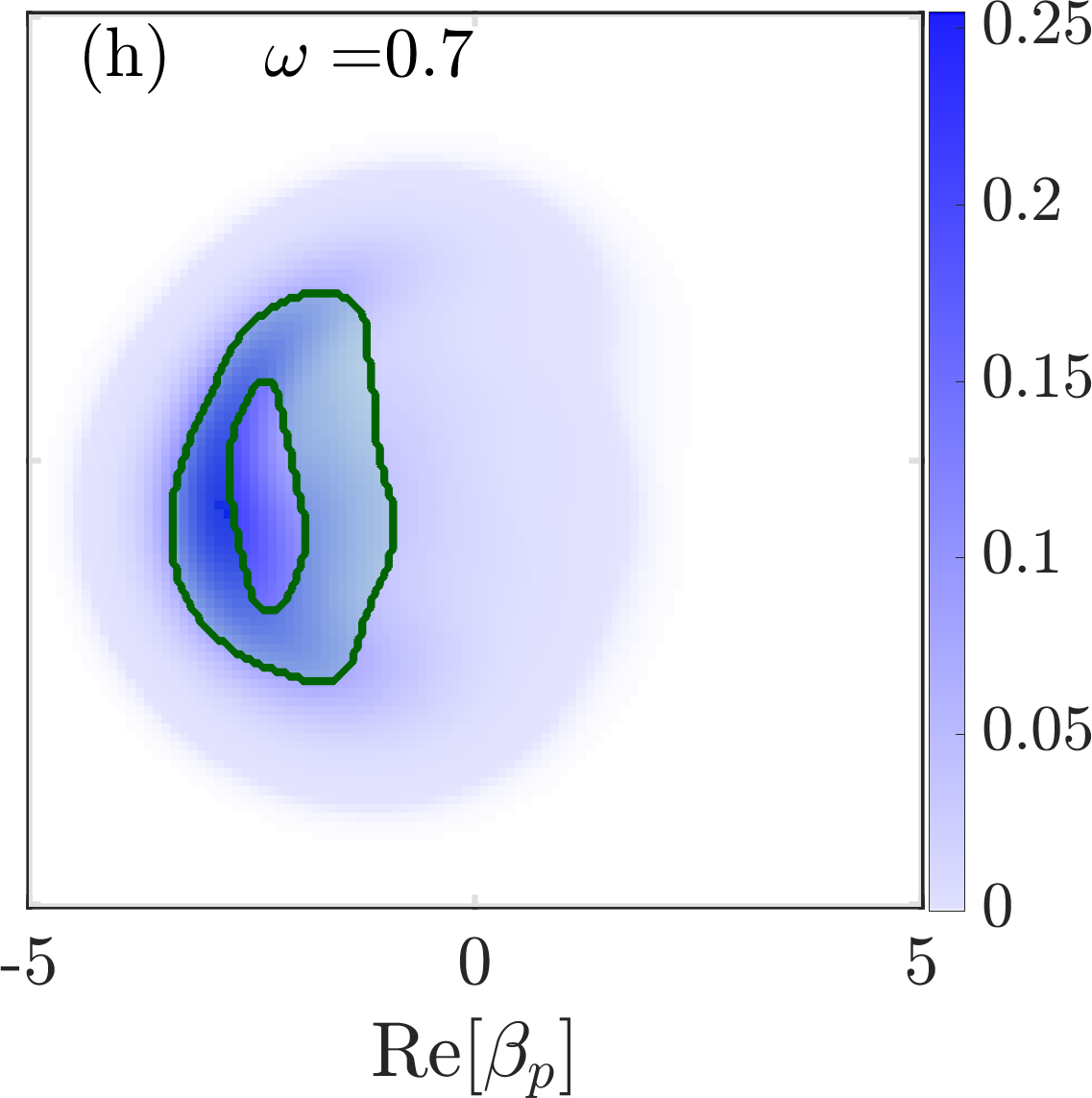} &
		\hspace{-1.8mm}\includegraphics[scale=0.2,valign=m]{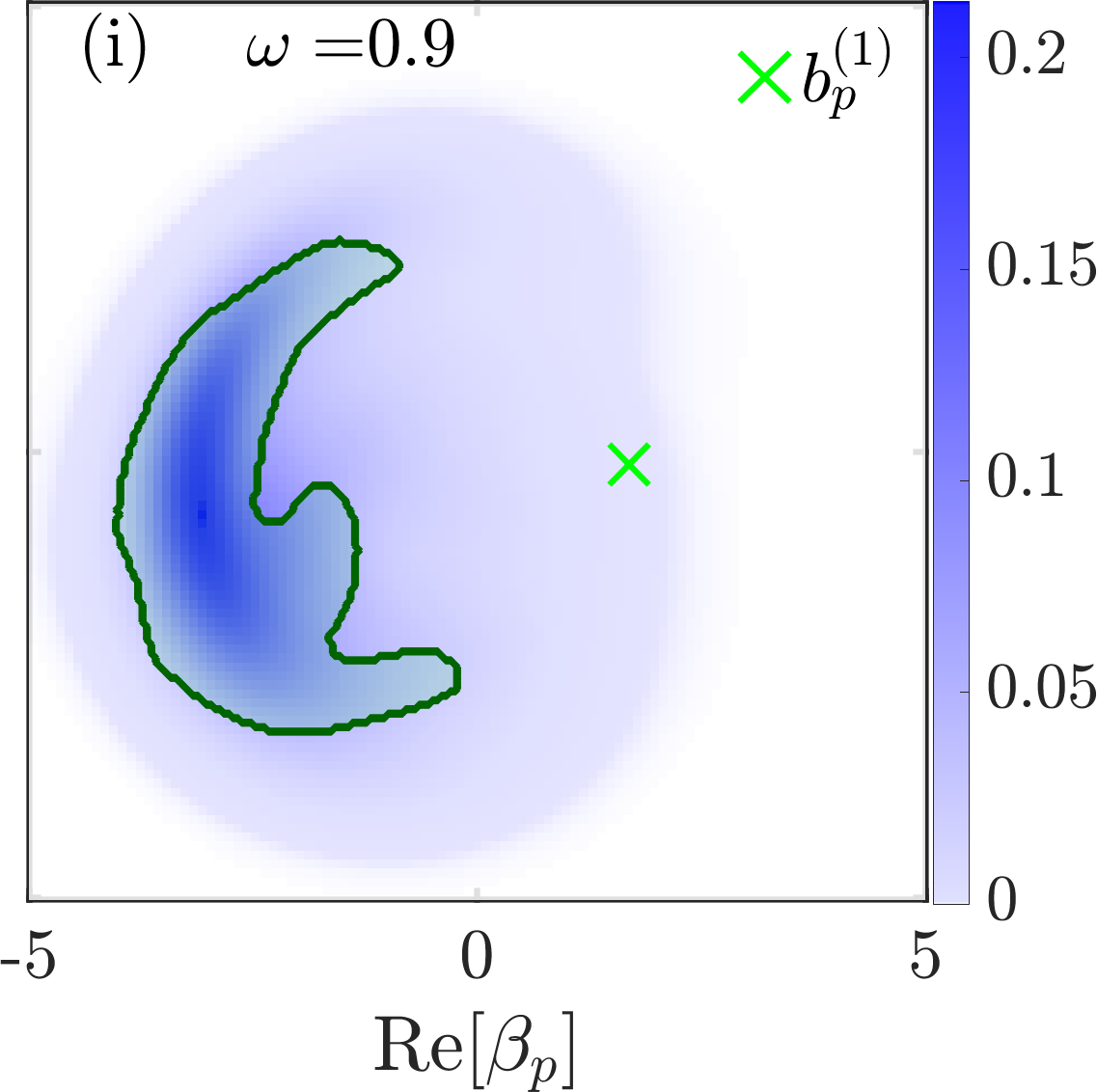} &
		\hspace{-1.8mm}\includegraphics[scale=0.2,valign=m]{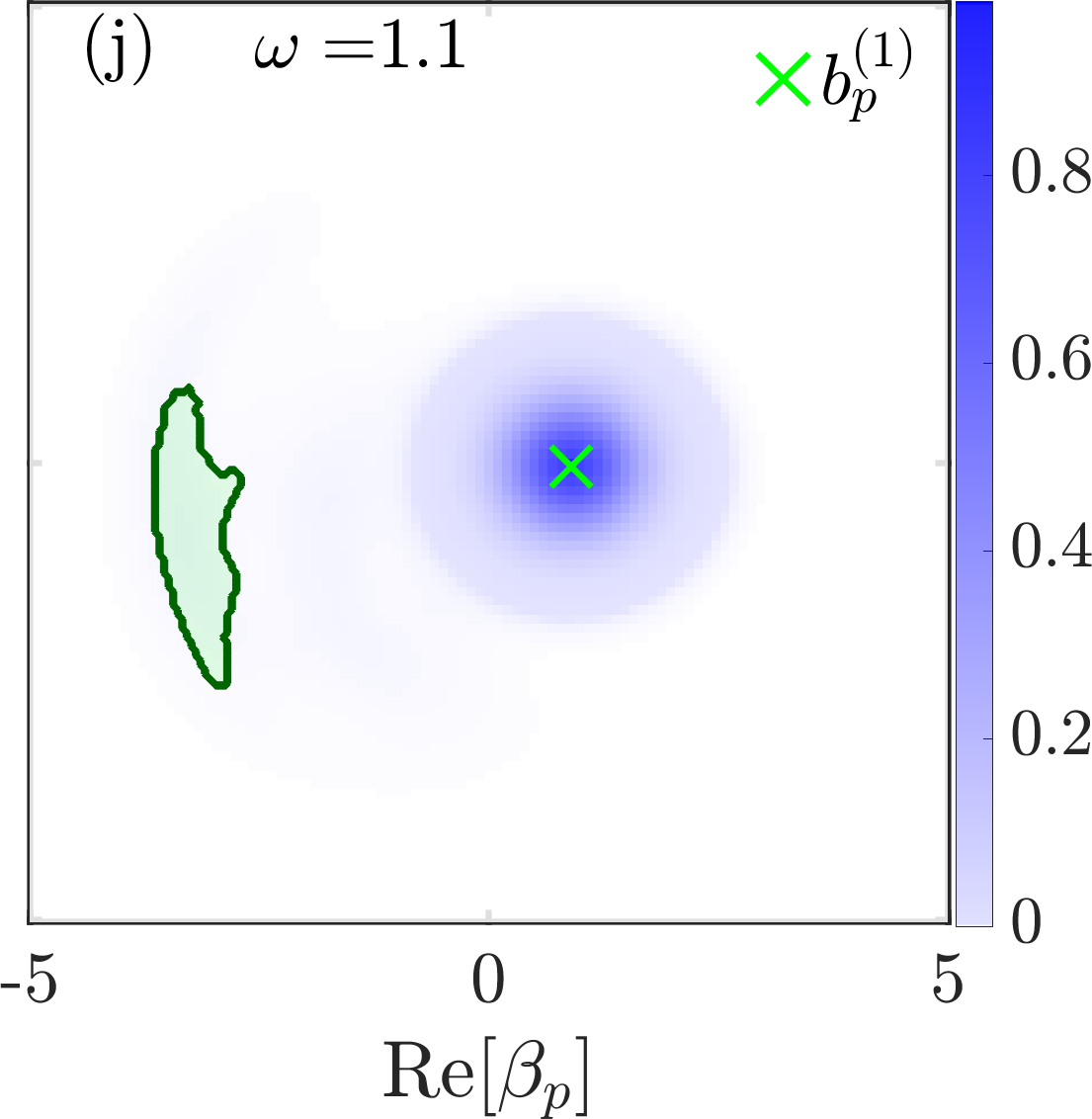} 	 
	\end{tabular}
	\caption{
		{\bf Wigner functions in the presence of a strange attractor.} The Wigner functions for frequencies below and within the strange attractor regime are presented for time $t_\text{m}=2000/J$. Panels (a)-(e) display the Wigner functions on a cut depicting the resonant ($\beta_p$) plane. Panels (f)-(j) display the Wigner function for the reduced density matrix associated with the same plane. The classical attractor is superimposed in green. The different fixed points are displayed using green markers.  Here $u=0.1$, $\gamma=0.05$, and $d=0.3$.}
	\label{fig:WignerFunction} 
	\vspace{-5mm}
\end{figure*}


\emph{Quantum approach to chaos.}--- We now turn to study the full quantum evolution of Eq.~\eqref{eq:lindblad} using the method of quantum trajectories \cite{molmer1993monte}, and compare with the classical equation of motion analysis. The quantum trajectories are visualized in Fig.~\ref{fig:quantum_trajectories} using a histogram, which depicts the formation and interplay of the various fixed points and their exploration by the quantum dynamics. In Fig.~\ref{fig:quantum_trajectories}a-c the trajectories explore the vicinity of the bright fixed point but as the frequency increases a larger portion of the chaotic basin is traversed quantum mechanically. Fig.~\ref{fig:quantum_trajectories}c-d depict a regime where the full classical chaotic basin is explored. Curiously, in Fig.~\ref{fig:quantum_trajectories}d the trajectories initially tend towards the newly nucleated dim fixed point but eventually flow past it and end up in the chaotic attractor. Fig.~\ref{fig:quantum_trajectories}e,f show the coexistence of two attractors, a chaotic one and a dim fixed point, as the weight of the former decreases with increasing frequency. 

The state the system attains at long times can be characterized by the Wigner function $W(\boldsymbol{\beta};t)$, of which we display only the two-dimensional cut with finite $\beta_p$ (and $\beta_k=0$ for $k\neq p$) in Fig.~\ref{fig:WignerFunction}a-e \cite{appmat}. When the system flows into a fixed point it generates a Wigner function that is localized around this fixed point, forming an approximate coherent state, see e.g. Fig.~\ref{fig:WignerFunction}e. In contrast, in the NFP regime, the Wigner function exhibits a pattern of islands of positive and negative values which aggregate around the SA, see Fig.~\ref{fig:WignerFunction}b-d. When projected onto the resonant plane, we introduce the Wigner function which characterizes the associated reduced density matrix, which becomes strictly positive, see Fig.~\ref{fig:WignerFunction}f-j. Remarkably, it admits the greatest support around the region of the classical SA in phase space, Fig.~\ref{fig:WignerFunction}h,i \cite{lee1993signatures}. 

Next, we consider photonic correlation measurements and weigh their relevance in unveiling the presence of quantum chaos in our system, starting with the standard photonic correlators $g^{(1)}$ and $g^{(2)}$, obtained by the  quantum trajectories  procedure \cite{appmat}.

\begin{figure*}[t]	
	\begin{minipage}[b]{0.49\linewidth}
		\includegraphics[width=0.97\linewidth]{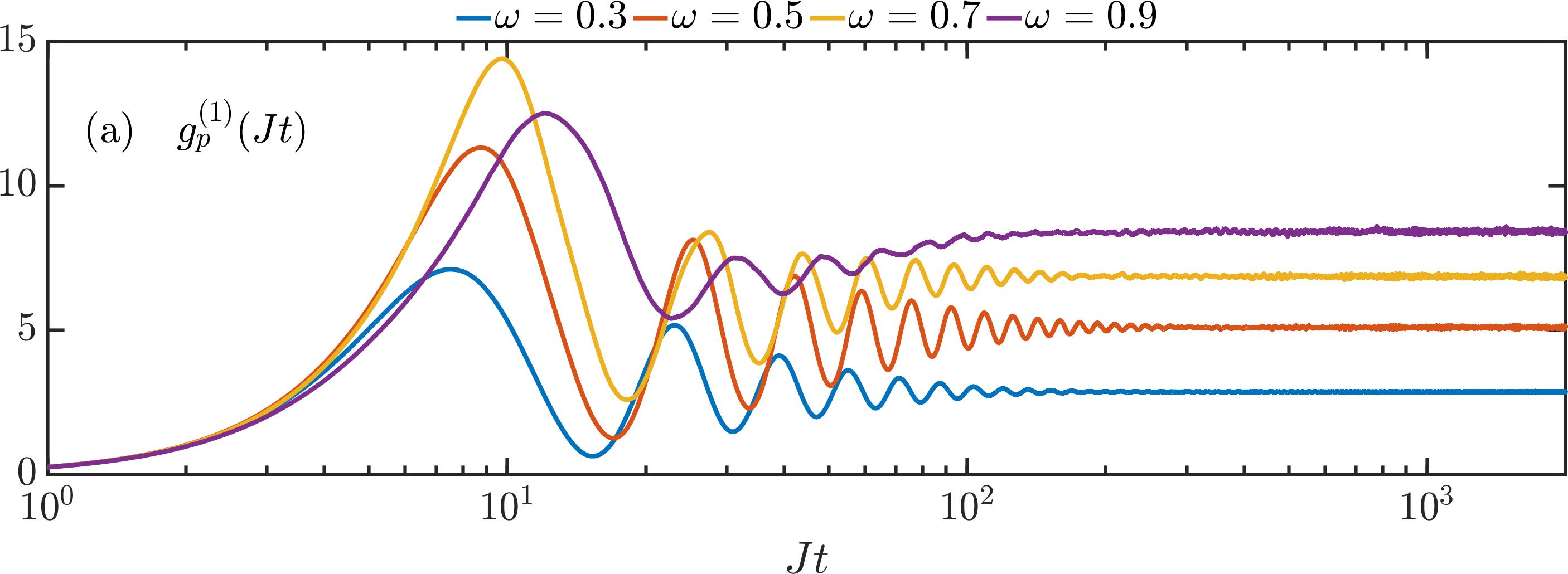}\vspace{0.1cm}\\
		\includegraphics[width=\linewidth]{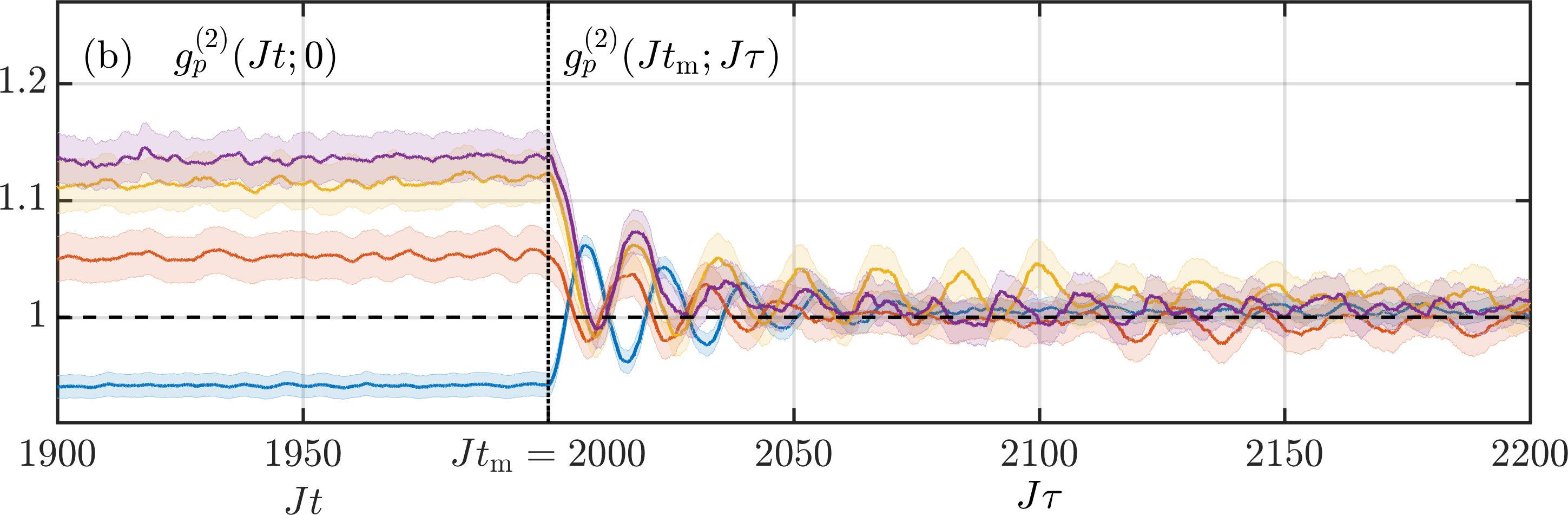}
	\end{minipage}
	\begin{minipage}[b]{0.49\linewidth}
		\includegraphics[width=\linewidth]{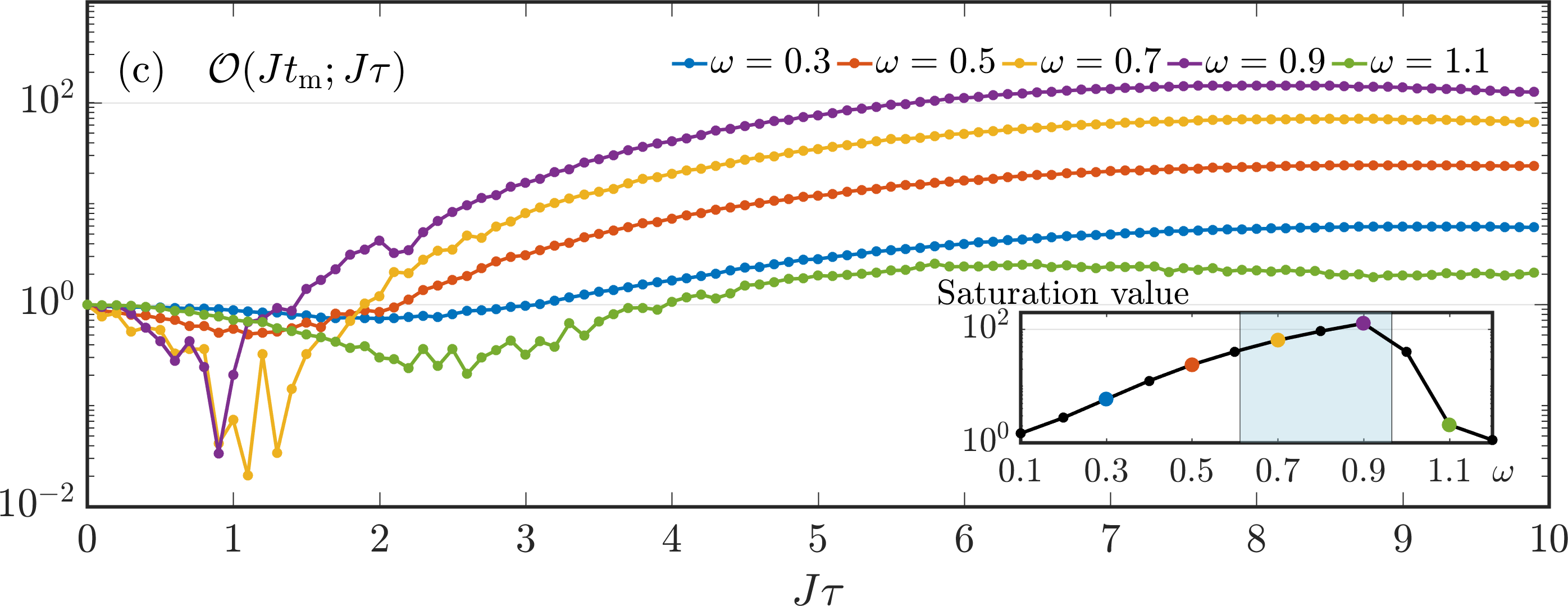}\vspace{0.1cm}
		\hspace*{0.3cm}\includegraphics[width=0.97\linewidth]{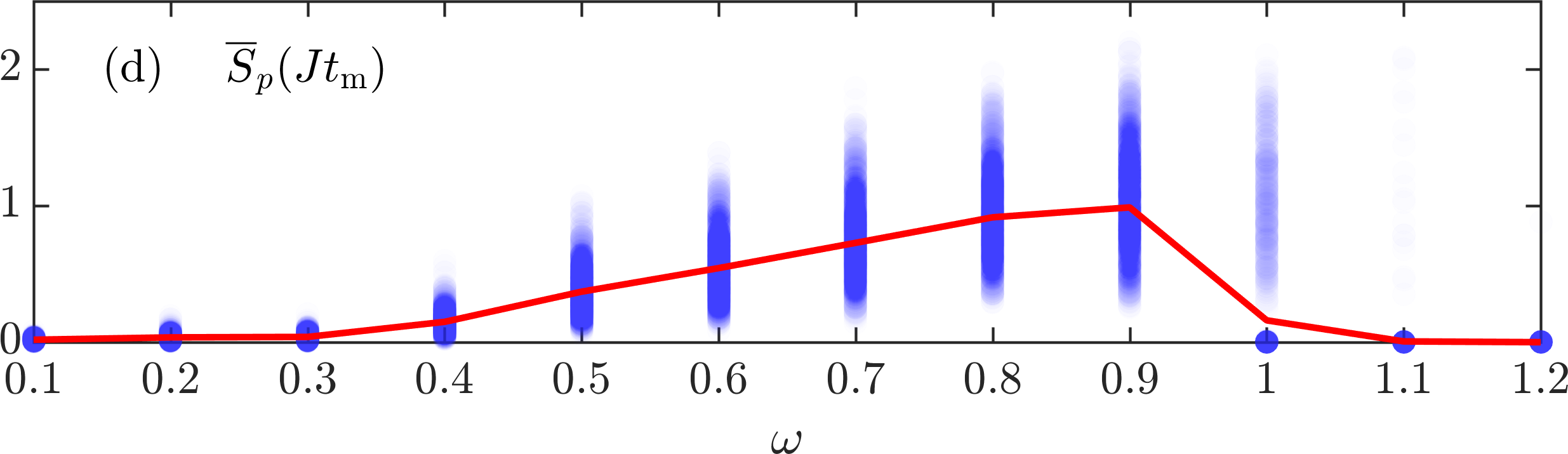}
	\end{minipage}
	\caption{
		{\bf Correlation functions and entanglement entropy.} Panels (a)-(b): the $g_p^{(1)}(Jt)$, $g_p^{(2)}(Jt;J\tau)$ photonic correlation functions, presented as function of time, for different frequencies. The shading denotes the standard deviation (not visible for $g_p^{(1)}$). In panel (b) the left side (with respect to the dashed vertical line) presents $g^{(2)}_p(Jt,0)$ as function of $Jt$ while the right side presents $g^{(2)}_p(Jt_m=2000,J\tau)$ as function of $J\tau$. Panel (c) displays the OTOC $\mathcal{O}(Jt_{\text{m}};J\tau)$, with the inset showing the saturation value for the different frequencies (black points are calculated but not displayed in the main panel). The blue shading in the inset denotes the NFP region of Fig.~\ref{fig:phase-diagram}b. Panel (d): the trajectory-averaged entanglement entropy $\overline{S}_p(J t_\text{m})$ (red line) as function of the drive frequency $\omega$, with the contribution of individual trajectories denoted by blue points. Here $u=0.1$, $\gamma=0.05$, $d=0.3$ and $t_\text{m}=2000/J$.}
	\label{fig:CorrelationFunctions} 
	\vspace{-5mm}
\end{figure*}

First, the correlator $g^{(1)}_p(t)=\langle \hat{b}^\dagger_p(t) \hat{b}_p(t)\rangle$ is associated with the photonic occupation and is presented in Fig.~\ref{fig:CorrelationFunctions}a. As the frequency increases, it is shown to reach a steady state value that progressively deviates from the value of the bright fixed point towards a mean value associated with the chaotic attractor. When the frequency further increases, the value of the $g^{(1)}$ correlator rapidly drops into the dim fixed point (not shown).

The correlator $g^{(2)}_p(t;\tau)=\langle \hat{b}^\dagger_p(t) \hat{b}^\dagger_p(t+\tau) \hat{b}_p(t+\tau) \hat{b}_p(t)\rangle/\langle \hat{b}^\dagger_p(t) \hat{b}_p(t)\rangle^2$ is presented in Fig.~\ref{fig:CorrelationFunctions}b and is shown to be sensitive to the presence of the SA as is demonstrated by the larger value it attains at $\tau=0$. This value is indicative of the increased fluctuations of the light generated by the chaotic phase space. In addition, the fluctuations of the quantum trajectories around the mean value could indicate the presence of the SA if the quantum trajectories are individually measured~\cite{murch2013observing}. 

The most prominent indicator of quantum chaos in our system turns out to be the OTOC \cite{larkin1969quasiclassical,maldacena2016bound}, defined by,
\begin{align}
	\mathcal{O}(t;\tau)=-\langle[\hat{Q}(t+\tau),\hat{P}(t)]^2\rangle,
\end{align}
where $\hat{Q}=(\hat{b}_p+\hat{b}_p^\dag)/\sqrt{2}$ and $\hat{P}=i(\hat{b}_p^\dag-\hat{b}_p)/\sqrt{2}$ are the dimensionless position and momentum operators associated with $\hat{b}_p$. The OTOC generalizes the concept of a Lyapunov exponent to the quantum regime, i.e., $\lim_{t\to\infty}\mathcal{O}(t;\tau)\sim e^{\lambda_q \tau}$ where $\lambda_q$ is the quantum Lyapunov exponent and $\tau$ is in some intermediate regime: for small enough $\tau$ the commutator is set by its equal-time value while for large enough $\tau$ it saturates~\cite{hashimoto2017out}. A version of the OTOC was measured in a trapped ion quantum magnet~\cite{garttner2017measuring}. In \cite{appmat} we detail a procedure to calculate it for a dissipative system.

The result, presented in Fig.~\ref{fig:CorrelationFunctions}c, exhibits the anticipated exponential increase within the chaotic regime. Remarkably, we find a non-monotonic behaviour of the saturation value of the OTOC as function of the drive frequency $\omega$ (see Fig.~\ref{fig:CorrelationFunctions}c, inset). It is interesting to note that a positive quantum Lyapunov exponent is detected for lower values of $\omega$ as compared to the classical phase diagram. At higher frequencies, once the frequency crosses into the classical dim fixed point regime, a dramatic reduction in the quantum Lyapunov exponent occurs. It therefore turns out that the quantum chaotic regime extends beyond the classical one into the lower frequency regime; see inset of Fig.~\ref{fig:CorrelationFunctions}c where the classical chaotic regime is depicted by blue shading. The higher frequency regime is marked by a coexistence of the dim and chaotic fixed points. Fig.~\ref{fig:CorrelationFunctions}d displays the entanglement entropy associated with the resonant mode calculated for individual trajectories, $S_p(t)=-\tr \left[\hat{\rho}_p(t) \log \hat{\rho}_p(t)\right]$, with $\hat{\rho}_p(t)=\tr_{k\neq p}|\psi(t)\rangle\langle\psi(t)|$ the reduced density matrix. For each trajectory $S_p(t)$ (blue points) is well defined, and the average over trajectories $\overline{S}_p(t)$ is displayed as a red line. It is interesting to note the spread of the quantum trajectories in the chaotic regime, which falls significantly once the system exits the chaotic regime. The similarity of the shape of the graph to the inset of panel (c) indicates relations between the entanglement entropy and the OTOC~\cite{lewis2019unifying}, that characterize the chaotic attractor.


\emph{Discussion.}--- We have demonstrated that the driven-dissipative Bose-Hubbard model with a small number of sites on a ring, once  chirally-driven, could lead to the generation of chaotic regimes in both the mean-field analysis and at the full quantum level. To characterize the quantum chaos, we have calculated the Wigner function on a cut through the 6-dimensional phase space and projected it onto the 2-dimensional resonant plane, where it is shown to spread over the approximate region of the classical chaotic phase space. Next, we extended the Monte-Carlo method to calculate the OTOC in driven-dissipative systems~\cite{appmat}, and demonstrated its sensitivity to the onset of quantum chaos in our model. Finally, we calculated the measurable correlators $g^{(1)}$ and $g^{(2)}$ and extracted the relevant signatures of the chaotic attractor. The results could prove relevant for photons in coupled cavity arrays and may lead to a study of chaos in such systems.


\begin{acknowledgments}

We thank D.~Cohen for discussions. This  research  was  funded  by  the  Israel  Innovation  Authority under the Kamin program as part of the QuantERA project InterPol, and by the Israel Science Foundation  under grant 1626/16.

\end{acknowledgments}


\appendix

\section{Mean field analysis}

To study the mean field fixed points we employ above the ansatz that only the single mode that shares its momentum with the drive is macroscopically occupied, i.e. $b_{k} = 0$ for $k\neq p$. The equation of motion for $b_{p}$ is given in this case by 
\begin{align}
i  \dot{b}_{p} = 
\frac{U }{M}|b_{p}|^2b_{p}
-\left[J \cos(p)
+\Omega+i\frac{\Gamma}{2}\right]b_{p} +\sqrt{M}D.
\end{align}
Denoting the steady state solution as $b_{p}(t\rightarrow \infty)=\sqrt{N_0}e^{-i\theta}$ we arrive at the following equation for the occupation of the condensate, $n_0=N_0/M$,
\begin{align}
\sqrt{n_0}\left[\cos (p)+\omega\right] = u n_0^{3/2} \pm  \sqrt{d^2-\frac{\gamma^2}{4}n_0},\label{eq:n0}
\end{align}
and its phase $\theta=\sin^{-1}(\gamma\sqrt{n_0}/2d)$. Eq.~\eqref{eq:n0} has three roots for $n_0$, which determine three fixed points $b_p^{(\nu)}=\sqrt{n_0^{(\nu)}}e^{-i\theta^{(\nu)}}$, with $\nu=1,2,3$. The stability of $b_p^{(\nu)}$ is then determined by perturbing $b_j=b^{(\nu)}_j+\delta b^{(\nu)}_j$, where $b^{(\nu)}_j= b^{(\nu)}_p e^{-ipx_j}/\sqrt{M}$ is the  steady state solution and $\delta b^{(\nu)}_j=e^{-\gamma s/2} e^{-i\theta^{(\nu)}}\delta \zeta_j$ with $s=J t$ the rescaled time. Taking into account only the leading terms in $\delta \zeta_j=e^{-ipx_j}(v_je^{-i\mu s}-u^*_je^{i\mu^* s})$, the equations of motion in the momentum basis $q$ prescribe 
\begin{equation}
\mu_q\left(\begin{array}{c}
u_{q}\\
v_{q}
\end{array}\right)
=
\mathcal L_q
\left(\begin{array}{c}
u_{q}\\
v_{q}
\end{array}\right).
\end{equation}
Here $\mathcal{L}_q= -\epsilon_0(q)+ i \sigma_y \epsilon_y+\sigma_z \epsilon_z(q)$,
where $\epsilon_0(q)=\sin(q) \sin(p)$, $\epsilon_y=n_0 u$, $\epsilon_z(q)=2 n_0 u  -\cos(q)\cos(p)-\omega$, $\sigma_i$ are the Pauli matrices and $\mu_q$ is the Bogoliubov dispersion, given by
$\mu_q=-\epsilon_0(q) \pm \sqrt{\epsilon_z^2(q)-\epsilon_y^2}$.
For a solution to be stable we require $\delta b_j$ to decay with time, thus stability is contingent upon $| \text{Im} \mu_q| \leq \gamma /2$ for all values of $q=\frac{2\pi}{M} m$, $m\in \{0,1,..,M-1\}$. This is an extension~\cite{carusotto2013quantum} of the more familiar case without loss ($\gamma=0$).

In the main text we describe the resulting phase diagram, which is presented in Fig.~\ref{fig:phase-diagram}a,b. Here we describe the phase diagram in terms of existence and stability of the fixed points. In region I (III) there exists only a single fixed point that is also stable: the dim (bright), $\nu=1$ ($\nu=3$), is the low (high) occupation fixed point. Whereas, in region II there exist three fixed points of which only the dim ($\nu=1$) one is stable. In addition there are the bistable regime (denoted by ``Bi'') in which both $\nu=1$ and $\nu=3$ are stable, and the regime where no fixed point is stable (denoted by ``NFP'') in which the system flows into the chaotic attractor, which is not revealed by the linear analysis above.

\section{Quantum analysis}

We calculate the Wigner function displayed in Fig.~\ref{fig:WignerFunction} using the formula
\begin{align}
W(\boldsymbol{\beta};t)=\int d\boldsymbol{\eta}\, d\boldsymbol{\eta}^* \rho_{\boldsymbol{\beta};\boldsymbol{\eta}}(t) e^{\frac{1}{2}\left(\boldsymbol{\eta}^*\cdot\boldsymbol{\beta}-\boldsymbol{\beta}^*\cdot\boldsymbol{\eta}\right)},
\end{align} 
where $\rho_{\boldsymbol{\beta};\boldsymbol{\eta}}(t)=\langle \boldsymbol{\beta}-\frac{\boldsymbol{\eta}}{2}|\hat{\rho}(t)|\boldsymbol{\beta}+\frac{\boldsymbol{\eta}}{2}\rangle$.
Here $\ket{\boldsymbol{\beta}}$ represents a product of momentum coherent states, i.e, $\ket{\boldsymbol \beta}= \bigotimes_k\ket{\beta_k},$  satisfying $\hat b_k \ket{\boldsymbol \beta} =\beta_k\ket{\boldsymbol \beta}$. Expressing the density matrix using momentum occupation states $\ket{\boldsymbol{n}}=\bigotimes_k \ket{n_k}$, we get the following transform
\begin{eqnarray}\label{eq:wigner}
W(\boldsymbol{\beta};t)=\sum_{\boldsymbol{n},\boldsymbol{n}'}\rho_{\boldsymbol{n},\boldsymbol{n}'}(t)\prod_{k} F_{n_k,n_k'}(2\beta_k),
\end{eqnarray}
where
\begin{align}
F_{n,n'}(\xi)=\frac{4e^{-|\xi|^2/2}\xi^{n'-n}}{\sqrt{n!n'!}}U(-n,n'-n+1,|\xi|^2),
\end{align}
with $U(a,b,c)$ denoting the confluent hypergeometric function. The density matrix $\rho_{\boldsymbol{n},\boldsymbol{n}'}(t)$ is then calculated employing $2000$ quantum trajectories. In the main text we considered two types of Wigner functions. The first was the full 6-dimensional Wigner function of Eq.~\eqref{eq:wigner} of which we displayed only the 2-dimensional cut representing the resonant plane ($\beta_k=0$ for $k\neq p$) at time $Jt_\text{m}$ (see Fig.\ref{fig:WignerFunction}a-e). Second, we consider the reduced density matrix relevant to the resonant momentum, i.e, the density matrix in Eq.~\eqref{eq:wigner} is replaced with $\hat \rho_{p}(t)= \tr_{k\neq p} \hat \rho(t)$ (see Fig.\ref{fig:WignerFunction}f-j).

Using the same quantum trajectory Monte-Carlo formalism, we calculate the different photonic correlation functions as we now describe. 

To calculate the $g^{(2)}_p(t;\tau)=\langle \hat{b}^\dagger_p(t) \hat{b}^\dagger_p(t+\tau) \hat{b}_p(t+\tau) \hat{b}_p(t)\rangle/\langle \hat{b}^\dagger_p(t) \hat{b}_p(t)\rangle^2$ correlator, we employ the usual method of first enforcing a quantum a jump at $t$ and then finding the evolution along $\tau$ following the same Monte-Carlo procedure.

Calculating the OTOC in a dissipative-driven system is more challenging. The OTOC can be expanded into a sum of four two-time correlators  with both forward and backward time evolutions. For each of these correlators, let $t_1$ be the time associated with the rightmost operator, and let $t_2$ be the other time. We proceed according to the following steps: (i), we evolve the wavefunction to $t_1$ (i.e. either $t$ or $t+\tau$); (ii), we apply the operators associated with $t_1$ on both the evolved ket and bra or, where this is not possible, we define four helper states following M\o{}lmer et al~\cite{molmer1993monte}; next, (iii), we apply either forward or backward evolution in time from $t_1$ to $t_2$, as necessary (i.e., forward evolution from $t_1=t$ to $t_2=t+\tau$ or backward evolution from $t_1=t+\tau$ to $t_2=t$). The latter is done according to $\hat{\mathcal{H}}\to-\hat{\mathcal{H}}$ but with the dissipative part of the Lindblad evolution remaining the same~\cite{tuziemski2019out}; finally, (iv), we find the expectation value of the operator or product of operators associated with time $t_2$.


\bibliography{DDBH.bib} 

\begin{thebibliography}{29}%
\makeatletter
\providecommand \@ifxundefined [1]{%
 \@ifx{#1\undefined}
}%
\providecommand \@ifnum [1]{%
 \ifnum #1\expandafter \@firstoftwo
 \else \expandafter \@secondoftwo
 \fi
}%
\providecommand \@ifx [1]{%
 \ifx #1\expandafter \@firstoftwo
 \else \expandafter \@secondoftwo
 \fi
}%
\providecommand \natexlab [1]{#1}%
\providecommand \enquote  [1]{``#1''}%
\providecommand \bibnamefont  [1]{#1}%
\providecommand \bibfnamefont [1]{#1}%
\providecommand \citenamefont [1]{#1}%
\providecommand \href@noop [0]{\@secondoftwo}%
\providecommand \href [0]{\begingroup \@sanitize@url \@href}%
\providecommand \@href[1]{\@@startlink{#1}\@@href}%
\providecommand \@@href[1]{\endgroup#1\@@endlink}%
\providecommand \@sanitize@url [0]{\catcode `\\12\catcode `\$12\catcode
  `\&12\catcode `\#12\catcode `\^12\catcode `\_12\catcode `\%12\relax}%
\providecommand \@@startlink[1]{}%
\providecommand \@@endlink[0]{}%
\providecommand \url  [0]{\begingroup\@sanitize@url \@url }%
\providecommand \@url [1]{\endgroup\@href {#1}{\urlprefix }}%
\providecommand \urlprefix  [0]{URL }%
\providecommand \Eprint [0]{\href }%
\providecommand \doibase [0]{https://doi.org/}%
\providecommand \selectlanguage [0]{\@gobble}%
\providecommand \bibinfo  [0]{\@secondoftwo}%
\providecommand \bibfield  [0]{\@secondoftwo}%
\providecommand \translation [1]{[#1]}%
\providecommand \BibitemOpen [0]{}%
\providecommand \bibitemStop [0]{}%
\providecommand \bibitemNoStop [0]{.\EOS\space}%
\providecommand \EOS [0]{\spacefactor3000\relax}%
\providecommand \BibitemShut  [1]{\csname bibitem#1\endcsname}%
\let\auto@bib@innerbib\@empty
\bibitem [{\citenamefont {Carusotto}\ and\ \citenamefont
  {Ciuti}(2013)}]{carusotto2013quantum}%
  \BibitemOpen
  \bibfield  {author} {\bibinfo {author} {\bibfnamefont {I.}~\bibnamefont
  {Carusotto}}\ and\ \bibinfo {author} {\bibfnamefont {C.}~\bibnamefont
  {Ciuti}},\ }\bibfield  {title} {\bibinfo {title} {Quantum fluids of light},\
  }\href {https://doi.org/10.1103/RevModPhys.85.299} {\bibfield  {journal}
  {\bibinfo  {journal} {Rev. Mod. Phys.}\ }\textbf {\bibinfo {volume} {85}},\
  \bibinfo {pages} {299} (\bibinfo {year} {2013})}\BibitemShut {NoStop}%
\bibitem [{\citenamefont {{Amo}}\ \emph {et~al.}(2009)\citenamefont {{Amo}},
  \citenamefont {{Lefr{\`e}re}}, \citenamefont {{Pigeon}}, \citenamefont
  {{Adrados}}, \citenamefont {{Ciuti}}, \citenamefont {{Carusotto}},
  \citenamefont {{Houdr{\'e}}}, \citenamefont {{Giacobino}},\ and\
  \citenamefont {{Bramati}}}]{amo2009superfluidity}%
  \BibitemOpen
  \bibfield  {author} {\bibinfo {author} {\bibfnamefont {A.}~\bibnamefont
  {{Amo}}}, \bibinfo {author} {\bibfnamefont {J.}~\bibnamefont
  {{Lefr{\`e}re}}}, \bibinfo {author} {\bibfnamefont {S.}~\bibnamefont
  {{Pigeon}}}, \bibinfo {author} {\bibfnamefont {C.}~\bibnamefont {{Adrados}}},
  \bibinfo {author} {\bibfnamefont {C.}~\bibnamefont {{Ciuti}}}, \bibinfo
  {author} {\bibfnamefont {I.}~\bibnamefont {{Carusotto}}}, \bibinfo {author}
  {\bibfnamefont {R.}~\bibnamefont {{Houdr{\'e}}}}, \bibinfo {author}
  {\bibfnamefont {E.}~\bibnamefont {{Giacobino}}},\ and\ \bibinfo {author}
  {\bibfnamefont {A.}~\bibnamefont {{Bramati}}},\ }\bibfield  {title} {\bibinfo
  {title} {{Superfluidity of polaritons in semiconductor microcavities}},\
  }\href {https://doi.org/10.1038/nphys1364} {\bibfield  {journal} {\bibinfo
  {journal} {Nature Physics}\ }\textbf {\bibinfo {volume} {5}},\ \bibinfo
  {pages} {805} (\bibinfo {year} {2009})}\BibitemShut {NoStop}%
\bibitem [{\citenamefont {Ozawa}\ \emph {et~al.}(2019)\citenamefont {Ozawa},
  \citenamefont {Price}, \citenamefont {Amo}, \citenamefont {Goldman},
  \citenamefont {Hafezi}, \citenamefont {Lu}, \citenamefont {Rechtsman},
  \citenamefont {Schuster}, \citenamefont {Simon}, \citenamefont {Zilberberg},\
  and\ \citenamefont {Carusotto}}]{ozawa2019topological}%
  \BibitemOpen
  \bibfield  {author} {\bibinfo {author} {\bibfnamefont {T.}~\bibnamefont
  {Ozawa}}, \bibinfo {author} {\bibfnamefont {H.~M.}\ \bibnamefont {Price}},
  \bibinfo {author} {\bibfnamefont {A.}~\bibnamefont {Amo}}, \bibinfo {author}
  {\bibfnamefont {N.}~\bibnamefont {Goldman}}, \bibinfo {author} {\bibfnamefont
  {M.}~\bibnamefont {Hafezi}}, \bibinfo {author} {\bibfnamefont
  {L.}~\bibnamefont {Lu}}, \bibinfo {author} {\bibfnamefont {M.~C.}\
  \bibnamefont {Rechtsman}}, \bibinfo {author} {\bibfnamefont {D.}~\bibnamefont
  {Schuster}}, \bibinfo {author} {\bibfnamefont {J.}~\bibnamefont {Simon}},
  \bibinfo {author} {\bibfnamefont {O.}~\bibnamefont {Zilberberg}},\ and\
  \bibinfo {author} {\bibfnamefont {I.}~\bibnamefont {Carusotto}},\ }\bibfield
  {title} {\bibinfo {title} {Topological photonics},\ }\href
  {https://doi.org/10.1103/RevModPhys.91.015006} {\bibfield  {journal}
  {\bibinfo  {journal} {Rev. Mod. Phys.}\ }\textbf {\bibinfo {volume} {91}},\
  \bibinfo {pages} {015006} (\bibinfo {year} {2019})}\BibitemShut {NoStop}%
\bibitem [{\citenamefont {{Deng}}\ \emph {et~al.}(2002)\citenamefont {{Deng}},
  \citenamefont {{Weihs}}, \citenamefont {{Santori}}, \citenamefont {{Bloch}},\
  and\ \citenamefont {{Yamamoto}}}]{deng2002condensation}%
  \BibitemOpen
  \bibfield  {author} {\bibinfo {author} {\bibfnamefont {H.}~\bibnamefont
  {{Deng}}}, \bibinfo {author} {\bibfnamefont {G.}~\bibnamefont {{Weihs}}},
  \bibinfo {author} {\bibfnamefont {C.}~\bibnamefont {{Santori}}}, \bibinfo
  {author} {\bibfnamefont {J.}~\bibnamefont {{Bloch}}},\ and\ \bibinfo {author}
  {\bibfnamefont {Y.}~\bibnamefont {{Yamamoto}}},\ }\bibfield  {title}
  {\bibinfo {title} {{Condensation of Semiconductor Microcavity Exciton
  Polaritons}},\ }\href {https://doi.org/10.1126/science.1074464} {\bibfield
  {journal} {\bibinfo  {journal} {Science}\ }\textbf {\bibinfo {volume}
  {298}},\ \bibinfo {pages} {199} (\bibinfo {year} {2002})}\BibitemShut
  {NoStop}%
\bibitem [{\citenamefont {{Schmidt}}\ and\ \citenamefont
  {{Koch}}(2013)}]{schmidt2013circuit}%
  \BibitemOpen
  \bibfield  {author} {\bibinfo {author} {\bibfnamefont {S.}~\bibnamefont
  {{Schmidt}}}\ and\ \bibinfo {author} {\bibfnamefont {J.}~\bibnamefont
  {{Koch}}},\ }\bibfield  {title} {\bibinfo {title} {{Circuit QED lattices:
  Towards quantum simulation with superconducting circuits}},\ }\href
  {https://doi.org/10.1002/andp.201200261} {\bibfield  {journal} {\bibinfo
  {journal} {Annalen der Physik}\ }\textbf {\bibinfo {volume} {525}},\ \bibinfo
  {pages} {395} (\bibinfo {year} {2013})}\BibitemShut {NoStop}%
\bibitem [{\citenamefont {Fitzpatrick}\ \emph {et~al.}(2017)\citenamefont
  {Fitzpatrick}, \citenamefont {Sundaresan}, \citenamefont {Li}, \citenamefont
  {Koch},\ and\ \citenamefont {Houck}}]{fitzpatrick2017observation}%
  \BibitemOpen
  \bibfield  {author} {\bibinfo {author} {\bibfnamefont {M.}~\bibnamefont
  {Fitzpatrick}}, \bibinfo {author} {\bibfnamefont {N.~M.}\ \bibnamefont
  {Sundaresan}}, \bibinfo {author} {\bibfnamefont {A.~C.~Y.}\ \bibnamefont
  {Li}}, \bibinfo {author} {\bibfnamefont {J.}~\bibnamefont {Koch}},\ and\
  \bibinfo {author} {\bibfnamefont {A.~A.}\ \bibnamefont {Houck}},\ }\bibfield
  {title} {\bibinfo {title} {Observation of a dissipative phase transition in a
  one-dimensional circuit qed lattice},\ }\href
  {https://doi.org/10.1103/PhysRevX.7.011016} {\bibfield  {journal} {\bibinfo
  {journal} {Phys. Rev. X}\ }\textbf {\bibinfo {volume} {7}},\ \bibinfo {pages}
  {011016} (\bibinfo {year} {2017})}\BibitemShut {NoStop}%
\bibitem [{\citenamefont {{Amo}}\ and\ \citenamefont
  {{Bloch}}(2016)}]{amo2016exciton}%
  \BibitemOpen
  \bibfield  {author} {\bibinfo {author} {\bibfnamefont {A.}~\bibnamefont
  {{Amo}}}\ and\ \bibinfo {author} {\bibfnamefont {J.}~\bibnamefont
  {{Bloch}}},\ }\bibfield  {title} {\bibinfo {title} {{Exciton-polaritons in
  lattices: A non-linear photonic simulator}},\ }\href
  {https://doi.org/10.1016/j.crhy.2016.08.007} {\bibfield  {journal} {\bibinfo
  {journal} {Comptes Rendus Physique}\ }\textbf {\bibinfo {volume} {17}},\
  \bibinfo {pages} {934} (\bibinfo {year} {2016})}\BibitemShut {NoStop}%
\bibitem [{\citenamefont {{Fink}}\ \emph {et~al.}(2018)\citenamefont {{Fink}},
  \citenamefont {{Schade}}, \citenamefont {{H{\"o}fling}}, \citenamefont
  {{Schneider}},\ and\ \citenamefont {{Imamoglu}}}]{fink2018signatures}%
  \BibitemOpen
  \bibfield  {author} {\bibinfo {author} {\bibfnamefont {T.}~\bibnamefont
  {{Fink}}}, \bibinfo {author} {\bibfnamefont {A.}~\bibnamefont {{Schade}}},
  \bibinfo {author} {\bibfnamefont {S.}~\bibnamefont {{H{\"o}fling}}}, \bibinfo
  {author} {\bibfnamefont {C.}~\bibnamefont {{Schneider}}},\ and\ \bibinfo
  {author} {\bibfnamefont {A.}~\bibnamefont {{Imamoglu}}},\ }\bibfield  {title}
  {\bibinfo {title} {{Signatures of a dissipative phase transition in photon
  correlation measurements}},\ }\href
  {https://doi.org/10.1038/s41567-017-0020-9} {\bibfield  {journal} {\bibinfo
  {journal} {Nature Physics}\ }\textbf {\bibinfo {volume} {14}},\ \bibinfo
  {pages} {365} (\bibinfo {year} {2018})}\BibitemShut {NoStop}%
\bibitem [{\citenamefont {{Abbarchi}}\ \emph {et~al.}(2013)\citenamefont
  {{Abbarchi}}, \citenamefont {{Amo}}, \citenamefont {{Sala}}, \citenamefont
  {{Solnyshkov}}, \citenamefont {{Flayac}}, \citenamefont {{Ferrier}},
  \citenamefont {{Sagnes}}, \citenamefont {{Galopin}}, \citenamefont
  {{Lema{\^\i}tre}}, \citenamefont {{Malpuech}},\ and\ \citenamefont
  {{Bloch}}}]{abbarchi2013macroscopic}%
  \BibitemOpen
  \bibfield  {author} {\bibinfo {author} {\bibfnamefont {M.}~\bibnamefont
  {{Abbarchi}}}, \bibinfo {author} {\bibfnamefont {A.}~\bibnamefont {{Amo}}},
  \bibinfo {author} {\bibfnamefont {V.~G.}\ \bibnamefont {{Sala}}}, \bibinfo
  {author} {\bibfnamefont {D.~D.}\ \bibnamefont {{Solnyshkov}}}, \bibinfo
  {author} {\bibfnamefont {H.}~\bibnamefont {{Flayac}}}, \bibinfo {author}
  {\bibfnamefont {L.}~\bibnamefont {{Ferrier}}}, \bibinfo {author}
  {\bibfnamefont {I.}~\bibnamefont {{Sagnes}}}, \bibinfo {author}
  {\bibfnamefont {E.}~\bibnamefont {{Galopin}}}, \bibinfo {author}
  {\bibfnamefont {A.}~\bibnamefont {{Lema{\^\i}tre}}}, \bibinfo {author}
  {\bibfnamefont {G.}~\bibnamefont {{Malpuech}}},\ and\ \bibinfo {author}
  {\bibfnamefont {J.}~\bibnamefont {{Bloch}}},\ }\bibfield  {title} {\bibinfo
  {title} {{Macroscopic quantum self-trapping and Josephson oscillations of
  exciton polaritons}},\ }\href {https://doi.org/10.1038/nphys2609} {\bibfield
  {journal} {\bibinfo  {journal} {Nature Physics}\ }\textbf {\bibinfo {volume}
  {9}},\ \bibinfo {pages} {275} (\bibinfo {year} {2013})}\BibitemShut {NoStop}%
\bibitem [{\citenamefont {Carlon~Zambon}\ \emph {et~al.}(2020)\citenamefont
  {Carlon~Zambon}, \citenamefont {Rodriguez}, \citenamefont {Lema\^{\i}tre},
  \citenamefont {Harouri}, \citenamefont {Le~Gratiet}, \citenamefont {Sagnes},
  \citenamefont {St-Jean}, \citenamefont {Ravets}, \citenamefont {Amo},\ and\
  \citenamefont {Bloch}}]{zambon2020parametric}%
  \BibitemOpen
  \bibfield  {author} {\bibinfo {author} {\bibfnamefont {N.}~\bibnamefont
  {Carlon~Zambon}}, \bibinfo {author} {\bibfnamefont {S.~R.~K.}\ \bibnamefont
  {Rodriguez}}, \bibinfo {author} {\bibfnamefont {A.}~\bibnamefont
  {Lema\^{\i}tre}}, \bibinfo {author} {\bibfnamefont {A.}~\bibnamefont
  {Harouri}}, \bibinfo {author} {\bibfnamefont {L.}~\bibnamefont {Le~Gratiet}},
  \bibinfo {author} {\bibfnamefont {I.}~\bibnamefont {Sagnes}}, \bibinfo
  {author} {\bibfnamefont {P.}~\bibnamefont {St-Jean}}, \bibinfo {author}
  {\bibfnamefont {S.}~\bibnamefont {Ravets}}, \bibinfo {author} {\bibfnamefont
  {A.}~\bibnamefont {Amo}},\ and\ \bibinfo {author} {\bibfnamefont
  {J.}~\bibnamefont {Bloch}},\ }\bibfield  {title} {\bibinfo {title}
  {Parametric instability in coupled nonlinear microcavities},\ }\href
  {https://doi.org/10.1103/PhysRevA.102.023526} {\bibfield  {journal} {\bibinfo
   {journal} {Phys. Rev. A}\ }\textbf {\bibinfo {volume} {102}},\ \bibinfo
  {pages} {023526} (\bibinfo {year} {2020})}\BibitemShut {NoStop}%
\bibitem [{\citenamefont {{Lled{\'o}}}\ \emph {et~al.}(2019)\citenamefont
  {{Lled{\'o}}}, \citenamefont {{Mavrogordatos}},\ and\ \citenamefont
  {{Szyma{\'n}ska}}}]{lledo2019driven}%
  \BibitemOpen
  \bibfield  {author} {\bibinfo {author} {\bibfnamefont {C.}~\bibnamefont
  {{Lled{\'o}}}}, \bibinfo {author} {\bibfnamefont {T.~K.}\ \bibnamefont
  {{Mavrogordatos}}},\ and\ \bibinfo {author} {\bibfnamefont {M.~H.}\
  \bibnamefont {{Szyma{\'n}ska}}},\ }\bibfield  {title} {\bibinfo {title}
  {{Driven Bose-Hubbard dimer under nonlocal dissipation: A bistable time
  crystal}},\ }\href {https://doi.org/10.1103/PhysRevB.100.054303} {\bibfield
  {journal} {\bibinfo  {journal} {\prb}\ }\textbf {\bibinfo {volume} {100}},\
  \bibinfo {eid} {054303} (\bibinfo {year} {2019})}\BibitemShut {NoStop}%
\bibitem [{\citenamefont {Sala}\ \emph {et~al.}(2015)\citenamefont {Sala},
  \citenamefont {Solnyshkov}, \citenamefont {Carusotto}, \citenamefont
  {Jacqmin}, \citenamefont {Lema\^{\i}tre}, \citenamefont
  {Ter\ifmmode~\mbox{\c{c}}\else \c{c}\fi{}as}, \citenamefont {Nalitov},
  \citenamefont {Abbarchi}, \citenamefont {Galopin}, \citenamefont {Sagnes},
  \citenamefont {Bloch}, \citenamefont {Malpuech},\ and\ \citenamefont
  {Amo}}]{sala2015spin}%
  \BibitemOpen
  \bibfield  {author} {\bibinfo {author} {\bibfnamefont {V.~G.}\ \bibnamefont
  {Sala}}, \bibinfo {author} {\bibfnamefont {D.~D.}\ \bibnamefont
  {Solnyshkov}}, \bibinfo {author} {\bibfnamefont {I.}~\bibnamefont
  {Carusotto}}, \bibinfo {author} {\bibfnamefont {T.}~\bibnamefont {Jacqmin}},
  \bibinfo {author} {\bibfnamefont {A.}~\bibnamefont {Lema\^{\i}tre}}, \bibinfo
  {author} {\bibfnamefont {H.}~\bibnamefont {Ter\ifmmode~\mbox{\c{c}}\else
  \c{c}\fi{}as}}, \bibinfo {author} {\bibfnamefont {A.}~\bibnamefont
  {Nalitov}}, \bibinfo {author} {\bibfnamefont {M.}~\bibnamefont {Abbarchi}},
  \bibinfo {author} {\bibfnamefont {E.}~\bibnamefont {Galopin}}, \bibinfo
  {author} {\bibfnamefont {I.}~\bibnamefont {Sagnes}}, \bibinfo {author}
  {\bibfnamefont {J.}~\bibnamefont {Bloch}}, \bibinfo {author} {\bibfnamefont
  {G.}~\bibnamefont {Malpuech}},\ and\ \bibinfo {author} {\bibfnamefont
  {A.}~\bibnamefont {Amo}},\ }\bibfield  {title} {\bibinfo {title} {Spin-orbit
  coupling for photons and polaritons in microstructures},\ }\href
  {https://doi.org/10.1103/PhysRevX.5.011034} {\bibfield  {journal} {\bibinfo
  {journal} {Phys. Rev. X}\ }\textbf {\bibinfo {volume} {5}},\ \bibinfo {pages}
  {011034} (\bibinfo {year} {2015})}\BibitemShut {NoStop}%
\bibitem [{\citenamefont {Solnyshkov}\ \emph {et~al.}(2009)\citenamefont
  {Solnyshkov}, \citenamefont {Johne}, \citenamefont {Shelykh},\ and\
  \citenamefont {Malpuech}}]{solnyshkov2009chaotic}%
  \BibitemOpen
  \bibfield  {author} {\bibinfo {author} {\bibfnamefont {D.~D.}\ \bibnamefont
  {Solnyshkov}}, \bibinfo {author} {\bibfnamefont {R.}~\bibnamefont {Johne}},
  \bibinfo {author} {\bibfnamefont {I.~A.}\ \bibnamefont {Shelykh}},\ and\
  \bibinfo {author} {\bibfnamefont {G.}~\bibnamefont {Malpuech}},\ }\bibfield
  {title} {\bibinfo {title} {Chaotic josephson oscillations of
  exciton-polaritons and their applications},\ }\href
  {https://doi.org/10.1103/PhysRevB.80.235303} {\bibfield  {journal} {\bibinfo
  {journal} {Phys. Rev. B}\ }\textbf {\bibinfo {volume} {80}},\ \bibinfo
  {pages} {235303} (\bibinfo {year} {2009})}\BibitemShut {NoStop}%
\bibitem [{\citenamefont {Gavrilov}(2016)}]{gavrilov2016towards}%
  \BibitemOpen
  \bibfield  {author} {\bibinfo {author} {\bibfnamefont {S.~S.}\ \bibnamefont
  {Gavrilov}},\ }\bibfield  {title} {\bibinfo {title} {Towards spin turbulence
  of light: Spontaneous disorder and chaos in cavity-polariton systems},\
  }\href {https://doi.org/10.1103/PhysRevB.94.195310} {\bibfield  {journal}
  {\bibinfo  {journal} {Phys. Rev. B}\ }\textbf {\bibinfo {volume} {94}},\
  \bibinfo {pages} {195310} (\bibinfo {year} {2016})}\BibitemShut {NoStop}%
\bibitem [{\citenamefont {Ruiz-S\'anchez}\ \emph {et~al.}(2020)\citenamefont
  {Ruiz-S\'anchez}, \citenamefont {Rechtman},\ and\ \citenamefont
  {Rubo}}]{sanchez2020autonomous}%
  \BibitemOpen
  \bibfield  {author} {\bibinfo {author} {\bibfnamefont {R.}~\bibnamefont
  {Ruiz-S\'anchez}}, \bibinfo {author} {\bibfnamefont {R.}~\bibnamefont
  {Rechtman}},\ and\ \bibinfo {author} {\bibfnamefont {Y.~G.}\ \bibnamefont
  {Rubo}},\ }\bibfield  {title} {\bibinfo {title} {Autonomous chaos of
  exciton-polariton condensates},\ }\href
  {https://doi.org/10.1103/PhysRevB.101.155305} {\bibfield  {journal} {\bibinfo
   {journal} {Phys. Rev. B}\ }\textbf {\bibinfo {volume} {101}},\ \bibinfo
  {pages} {155305} (\bibinfo {year} {2020})}\BibitemShut {NoStop}%
\bibitem [{\citenamefont {Dholakia}\ \emph {et~al.}(1996)\citenamefont
  {Dholakia}, \citenamefont {Simpson}, \citenamefont {Padgett},\ and\
  \citenamefont {Allen}}]{dholakia1996second}%
  \BibitemOpen
  \bibfield  {author} {\bibinfo {author} {\bibfnamefont {K.}~\bibnamefont
  {Dholakia}}, \bibinfo {author} {\bibfnamefont {N.~B.}\ \bibnamefont
  {Simpson}}, \bibinfo {author} {\bibfnamefont {M.~J.}\ \bibnamefont
  {Padgett}},\ and\ \bibinfo {author} {\bibfnamefont {L.}~\bibnamefont
  {Allen}},\ }\bibfield  {title} {\bibinfo {title} {Second-harmonic generation
  and the orbital angular momentum of light},\ }\href
  {https://doi.org/10.1103/PhysRevA.54.R3742} {\bibfield  {journal} {\bibinfo
  {journal} {Phys. Rev. A}\ }\textbf {\bibinfo {volume} {54}},\ \bibinfo
  {pages} {R3742} (\bibinfo {year} {1996})}\BibitemShut {NoStop}%
\bibitem [{\citenamefont {Martinelli}\ \emph {et~al.}(2004)\citenamefont
  {Martinelli}, \citenamefont {Huguenin}, \citenamefont {Nussenzveig},\ and\
  \citenamefont {Khoury}}]{martinelli2004orbital}%
  \BibitemOpen
  \bibfield  {author} {\bibinfo {author} {\bibfnamefont {M.}~\bibnamefont
  {Martinelli}}, \bibinfo {author} {\bibfnamefont {J.~A.~O.}\ \bibnamefont
  {Huguenin}}, \bibinfo {author} {\bibfnamefont {P.}~\bibnamefont
  {Nussenzveig}},\ and\ \bibinfo {author} {\bibfnamefont {A.~Z.}\ \bibnamefont
  {Khoury}},\ }\bibfield  {title} {\bibinfo {title} {Orbital angular momentum
  exchange in an optical parametric oscillator},\ }\href
  {https://doi.org/10.1103/PhysRevA.70.013812} {\bibfield  {journal} {\bibinfo
  {journal} {Phys. Rev. A}\ }\textbf {\bibinfo {volume} {70}},\ \bibinfo
  {pages} {013812} (\bibinfo {year} {2004})}\BibitemShut {NoStop}%
\bibitem [{\citenamefont {Zambon}\ \emph {et~al.}(2019)\citenamefont {Zambon},
  \citenamefont {St-Jean}, \citenamefont {Lema\^{i}tre}, \citenamefont
  {Harouri}, \citenamefont {Gratiet}, \citenamefont {Sagnes}, \citenamefont
  {Ravets}, \citenamefont {Amo},\ and\ \citenamefont
  {Bloch}}]{zambon2019circular}%
  \BibitemOpen
  \bibfield  {author} {\bibinfo {author} {\bibfnamefont {N.~C.}\ \bibnamefont
  {Zambon}}, \bibinfo {author} {\bibfnamefont {P.}~\bibnamefont {St-Jean}},
  \bibinfo {author} {\bibfnamefont {A.}~\bibnamefont {Lema\^{i}tre}}, \bibinfo
  {author} {\bibfnamefont {A.}~\bibnamefont {Harouri}}, \bibinfo {author}
  {\bibfnamefont {L.~L.}\ \bibnamefont {Gratiet}}, \bibinfo {author}
  {\bibfnamefont {I.}~\bibnamefont {Sagnes}}, \bibinfo {author} {\bibfnamefont
  {S.}~\bibnamefont {Ravets}}, \bibinfo {author} {\bibfnamefont
  {A.}~\bibnamefont {Amo}},\ and\ \bibinfo {author} {\bibfnamefont
  {J.}~\bibnamefont {Bloch}},\ }\bibfield  {title} {\bibinfo {title} {Orbital
  angular momentum bistability in a microlaser},\ }\href
  {https://doi.org/10.1364/OL.44.004531} {\bibfield  {journal} {\bibinfo
  {journal} {Opt. Lett.}\ }\textbf {\bibinfo {volume} {44}},\ \bibinfo {pages}
  {4531} (\bibinfo {year} {2019})}\BibitemShut {NoStop}%
\bibitem [{\citenamefont {{M{\o}lmer}}\ \emph {et~al.}(1993)\citenamefont
  {{M{\o}lmer}}, \citenamefont {{Castin}},\ and\ \citenamefont
  {{Dalibard}}}]{molmer1993monte}%
  \BibitemOpen
  \bibfield  {author} {\bibinfo {author} {\bibfnamefont {K.}~\bibnamefont
  {{M{\o}lmer}}}, \bibinfo {author} {\bibfnamefont {Y.}~\bibnamefont
  {{Castin}}},\ and\ \bibinfo {author} {\bibfnamefont {J.}~\bibnamefont
  {{Dalibard}}},\ }\bibfield  {title} {\bibinfo {title} {{Monte Carlo
  wave-function method in quantum optics}},\ }\href
  {https://doi.org/10.1364/JOSAB.10.000524} {\bibfield  {journal} {\bibinfo
  {journal} {Journal of the Optical Society of America B Optical Physics}\
  }\textbf {\bibinfo {volume} {10}},\ \bibinfo {pages} {524} (\bibinfo {year}
  {1993})}\BibitemShut {NoStop}%
\bibitem [{app()}]{appmat}%
  \BibitemOpen
  \href@noop {} {\bibinfo {title} {See appendix for additional details relating
  to the calculations}}\BibitemShut {NoStop}%
\bibitem [{\citenamefont {Kolovsky}(2020)}]{kolovsky2020bistability}%
  \BibitemOpen
  \bibfield  {author} {\bibinfo {author} {\bibfnamefont {A.~R.}\ \bibnamefont
  {Kolovsky}},\ }\bibfield  {title} {\bibinfo {title} {Bistability in the
  dissipative quantum systems i: Damped and driven nonlinear oscillator},\
  }\href {https://arxiv.org/abs/2002.11373} {\bibfield  {journal} {\bibinfo
  {journal} {arXiv preprint arXiv:2002.11373}\ } (\bibinfo {year}
  {2020})}\BibitemShut {NoStop}%
\bibitem [{\citenamefont {Lee}\ and\ \citenamefont
  {Feit}(1993)}]{lee1993signatures}%
  \BibitemOpen
  \bibfield  {author} {\bibinfo {author} {\bibfnamefont {S.}~\bibnamefont
  {Lee}}\ and\ \bibinfo {author} {\bibfnamefont {M.}~\bibnamefont {Feit}},\
  }\bibfield  {title} {\bibinfo {title} {Signatures of quantum chaos in wigner
  and husimi representations},\ }\href
  {https://journals.aps.org/pre/abstract/10.1103/PhysRevE.47.4552} {\bibfield
  {journal} {\bibinfo  {journal} {Physical Review E}\ }\textbf {\bibinfo
  {volume} {47}},\ \bibinfo {pages} {4552} (\bibinfo {year}
  {1993})}\BibitemShut {NoStop}%
\bibitem [{\citenamefont {Murch}\ \emph {et~al.}(2013)\citenamefont {Murch},
  \citenamefont {Weber}, \citenamefont {Macklin},\ and\ \citenamefont
  {Siddiqi}}]{murch2013observing}%
  \BibitemOpen
  \bibfield  {author} {\bibinfo {author} {\bibfnamefont {K.}~\bibnamefont
  {Murch}}, \bibinfo {author} {\bibfnamefont {S.}~\bibnamefont {Weber}},
  \bibinfo {author} {\bibfnamefont {C.}~\bibnamefont {Macklin}},\ and\ \bibinfo
  {author} {\bibfnamefont {I.}~\bibnamefont {Siddiqi}},\ }\bibfield  {title}
  {\bibinfo {title} {Observing single quantum trajectories of a superconducting
  quantum bit},\ }\href {https://www.nature.com/articles/nature12539}
  {\bibfield  {journal} {\bibinfo  {journal} {Nature}\ }\textbf {\bibinfo
  {volume} {502}},\ \bibinfo {pages} {211} (\bibinfo {year}
  {2013})}\BibitemShut {NoStop}%
\bibitem [{\citenamefont {{Larkin}}\ and\ \citenamefont
  {{Ovchinnikov}}(1969)}]{larkin1969quasiclassical}%
  \BibitemOpen
  \bibfield  {author} {\bibinfo {author} {\bibfnamefont {A.~I.}\ \bibnamefont
  {{Larkin}}}\ and\ \bibinfo {author} {\bibfnamefont {Y.~N.}\ \bibnamefont
  {{Ovchinnikov}}},\ }\bibfield  {title} {\bibinfo {title} {{Quasiclassical
  Method in the Theory of Superconductivity}},\ }\href@noop {} {\bibfield
  {journal} {\bibinfo  {journal} {Soviet Journal of Experimental and
  Theoretical Physics}\ }\textbf {\bibinfo {volume} {28}},\ \bibinfo {pages}
  {1200} (\bibinfo {year} {1969})}\BibitemShut {NoStop}%
\bibitem [{\citenamefont {{Maldacena}}\ \emph {et~al.}(2016)\citenamefont
  {{Maldacena}}, \citenamefont {{Shenker}},\ and\ \citenamefont
  {{Stanford}}}]{maldacena2016bound}%
  \BibitemOpen
  \bibfield  {author} {\bibinfo {author} {\bibfnamefont {J.}~\bibnamefont
  {{Maldacena}}}, \bibinfo {author} {\bibfnamefont {S.~H.}\ \bibnamefont
  {{Shenker}}},\ and\ \bibinfo {author} {\bibfnamefont {D.}~\bibnamefont
  {{Stanford}}},\ }\bibfield  {title} {\bibinfo {title} {{A bound on chaos}},\
  }\href {https://doi.org/10.1007/JHEP08(2016)106} {\bibfield  {journal}
  {\bibinfo  {journal} {Journal of High Energy Physics}\ }\textbf {\bibinfo
  {volume} {2016}},\ \bibinfo {eid} {106} (\bibinfo {year} {2016})}\BibitemShut
  {NoStop}%
\bibitem [{\citenamefont {Hashimoto}\ \emph {et~al.}(2017)\citenamefont
  {Hashimoto}, \citenamefont {Murata},\ and\ \citenamefont
  {Yoshii}}]{hashimoto2017out}%
  \BibitemOpen
  \bibfield  {author} {\bibinfo {author} {\bibfnamefont {K.}~\bibnamefont
  {Hashimoto}}, \bibinfo {author} {\bibfnamefont {K.}~\bibnamefont {Murata}},\
  and\ \bibinfo {author} {\bibfnamefont {R.}~\bibnamefont {Yoshii}},\
  }\bibfield  {title} {\bibinfo {title} {Out-of-time-order correlators in
  quantum mechanics},\ }\href
  {https://link.springer.com/article/10.1007/JHEP10(2017)138} {\bibfield
  {journal} {\bibinfo  {journal} {Journal of High Energy Physics}\ }\textbf
  {\bibinfo {volume} {2017}},\ \bibinfo {pages} {1} (\bibinfo {year}
  {2017})}\BibitemShut {NoStop}%
\bibitem [{\citenamefont {{G{\"a}rttner}}\ \emph {et~al.}(2017)\citenamefont
  {{G{\"a}rttner}}, \citenamefont {{Bohnet}}, \citenamefont {{Safavi-Naini}},
  \citenamefont {{Wall}}, \citenamefont {{Bollinger}},\ and\ \citenamefont
  {{Rey}}}]{garttner2017measuring}%
  \BibitemOpen
  \bibfield  {author} {\bibinfo {author} {\bibfnamefont {M.}~\bibnamefont
  {{G{\"a}rttner}}}, \bibinfo {author} {\bibfnamefont {J.~G.}\ \bibnamefont
  {{Bohnet}}}, \bibinfo {author} {\bibfnamefont {A.}~\bibnamefont
  {{Safavi-Naini}}}, \bibinfo {author} {\bibfnamefont {M.~L.}\ \bibnamefont
  {{Wall}}}, \bibinfo {author} {\bibfnamefont {J.~J.}\ \bibnamefont
  {{Bollinger}}},\ and\ \bibinfo {author} {\bibfnamefont {A.~M.}\ \bibnamefont
  {{Rey}}},\ }\bibfield  {title} {\bibinfo {title} {{Measuring
  out-of-time-order correlations and multiple quantum spectra in a trapped-ion
  quantum magnet}},\ }\href {https://doi.org/10.1038/nphys4119} {\bibfield
  {journal} {\bibinfo  {journal} {Nature Physics}\ }\textbf {\bibinfo {volume}
  {13}},\ \bibinfo {pages} {781} (\bibinfo {year} {2017})}\BibitemShut
  {NoStop}%
\bibitem [{\citenamefont {Lewis-Swan}\ \emph {et~al.}(2019)\citenamefont
  {Lewis-Swan}, \citenamefont {Safavi-Naini}, \citenamefont {Bollinger},\ and\
  \citenamefont {Rey}}]{lewis2019unifying}%
  \BibitemOpen
  \bibfield  {author} {\bibinfo {author} {\bibfnamefont {R.}~\bibnamefont
  {Lewis-Swan}}, \bibinfo {author} {\bibfnamefont {A.}~\bibnamefont
  {Safavi-Naini}}, \bibinfo {author} {\bibfnamefont {J.~J.}\ \bibnamefont
  {Bollinger}},\ and\ \bibinfo {author} {\bibfnamefont {A.~M.}\ \bibnamefont
  {Rey}},\ }\bibfield  {title} {\bibinfo {title} {Unifying scrambling,
  thermalization and entanglement through measurement of fidelity
  out-of-time-order correlators in the dicke model},\ }\href
  {https://www.nature.com/articles/s41467-019-09436-y} {\bibfield  {journal}
  {\bibinfo  {journal} {Nature communications}\ }\textbf {\bibinfo {volume}
  {10}},\ \bibinfo {pages} {1} (\bibinfo {year} {2019})}\BibitemShut {NoStop}%
\bibitem [{\citenamefont {Tuziemski}(2019)}]{tuziemski2019out}%
  \BibitemOpen
  \bibfield  {author} {\bibinfo {author} {\bibfnamefont {J.}~\bibnamefont
  {Tuziemski}},\ }\bibfield  {title} {\bibinfo {title} {Out-of-time-ordered
  correlation functions in open systems: A feynman-vernon influence functional
  approach},\ }\href
  {https://journals.aps.org/pra/abstract/10.1103/PhysRevA.100.062106}
  {\bibfield  {journal} {\bibinfo  {journal} {Physical Review A}\ }\textbf
  {\bibinfo {volume} {100}},\ \bibinfo {pages} {062106} (\bibinfo {year}
  {2019})}\BibitemShut {NoStop}%
\end{thebibliography}%

\end{document}